\documentclass[sort&compress]{elsarticle}
\usepackage{graphicx} 
\usepackage{amsmath}
\usepackage{amssymb}
\usepackage{amsfonts}
\usepackage{natbib}
\usepackage{IEEEtrantools}
\usepackage{appendix}
\newcommand{\ai}{\emph{ab initio}}
\newcommand{\EinA}{Einstein \emph{A} coefficients}
\newcommand{\etal}{\emph{et al.}}
\newcommand{\Brooke}{\textbf{Brooke16}}
\newcommand{\Noll}{\textbf{Noll20}}
\usepackage{color}

\title{Three-states model for calculating the $X$-$X$ rovibrational transition intensities in hydroxyl radical}
\author{V.~G.~Ushakov$^a$}
\author{A.~Yu.~Ermilov$^b$}
\author{E.~S.~Medvedev$^a$}

\date{October 15, 2024}

\address{$^a$Federal Research Center of Problems of Chemical Physics and Medicinal Chemistry (former Institute of Problems of Chemical Physics), Russian Academy of Sciences, 142432 Chernogolovka, Russian Federation}
\address{$^b$M.~V.~Lomonosov Moscow State University, Russian Federation}

\begin{document}

\begin{abstract}
    The best available line list of OH [Brooke \etal\ JQSRT, 168 (2016) 142] contains the high-quality line frequencies, yet the line intensities need refinement because the model function used to interpolate the RKR potential and to extrapolate it into the repulsion region was not analytic [Medvedev \etal\ Mol. Phys. doi: 10.1080/00268976.2024.2395439], and also because the coupling between the ground $X^2\Pi$ and first excited $A^2\Sigma^+$ electronic states was treated by the perturbation theory. In this paper, we performed \ai\ calculations of all necessary molecular functions at $r=0.4$-8.0 bohr, and then we construct fully analytic model functions entering the Hamiltonian. The model functions were fitted to both the \ai\ data and the available experimental data on the line positions and energy levels, the relative line intensities, and the transition dipole moments derived from the measured permanent dipoles. The system of three coupled Schr\"odinger equations for two multiplet components of the $X$ state plus the $A$ state was solved to calculate the energy levels and the line intensities. The new set of the \EinA\ permits to decrease the scatter of the logarithmic populations of the ro-vibrational levels derived from the observed radiation fluxes [Noll \etal\ Atmos. Chem. Phys. 20 (2020) 5269], to achieve better agreement with the measured relative intensities, and to obtain significant differences in the intensities of the $\Lambda$ doublets for large $v$ and $J$ as observed by Noll \emph{et al.} The $X$-$A$ coupling fully modifies the Q-line intensities at high $J$ by removing the well-known $J^{-2}$ dependence. A new line list is constructed where the transition frequencies are from Brooke \etal\ and the \EinA\ are from the present study. However, not all the problems with the intensities were resolved, presumably due to the neglect of the interaction with the $^4\Sigma^-,^2\Sigma^-$ and $^4\Pi$ repulsive electronic terms. 
\end{abstract}

\maketitle

\section{Introduction}

The importance of OH for studies of the terrestrial atmosphere is well described in the recent study by Noll \emph{et al.} \cite{Noll20} (\Noll) and in the HITRAN20 \cite{Gordon22}. 

The reduction of ozone by atomic hydrogen in the upper mesosphere creates hydroxyl at excited vibrational levels $v^\prime=$ 6-9 and excited rotational states \cite{Franzen19}. Noll \emph{et al.} measured 723 emission lines from the upper levels $v^\prime=3$--9 belonging to vibrational bands $\Delta v=3$--6. Using six available sets of the calculated Einstein $A$ coefficients \cite{Loo08,Turnbull89,Langhoff86,Goldman98,Brooke16,Mies74}, those of \Brooke\ \cite{Brooke16},  \textbf{Mies74} \cite{Mies74}, and HITRAN12 (based on \cite{Goldman98}) among them, they derived six sets of the logarithmic populations of the OH ro-vibrational states,
\begin{equation}
    y_{i^\prime}=\ln\left(\frac{I_{i^\prime i^{\prime\prime}}}{A_{i^\prime i^{\prime\prime}} g_{i^\prime}} \right),
    \label{y}
\end{equation}
where $I_{i^\prime i^{\prime\prime}}$ is the line intensity in rayleigh (1 R = $10^{10}$ photons$\cdot$m$^{-2}\cdot$s$^{-1}$), $A_{i^\prime i^{\prime\prime}}$ the \EinA\ in s$^{-1}$, and $g_{i^\prime}$ the total degeneracy of the upper level.

The wealth and high precision of the observational data allowed Noll \emph{et al.} to perform a detailed analysis in terms of the population differences, $\Delta y_{i^\prime}$, derived from the line pairs emitted from a common upper level $i^\prime$. Since the uncertainty of the flux measurement in Ref. \cite{Noll20} is estimated to be about 10\%, we assume that the expected values of $\Delta y_{i^\prime}$ are within $\pm0.15$. In fact, however, they vary from $+0.2$ to $-1.5$ (see Fig. 6 in \textbf{Noll20}),\footnote{In the supplementary file ``OHpop\_Acoeff\_basic" to \textbf{Noll20}, even higher values, up to 3.5 in the 6-2 band, can be found.} which testifies the insufficient quality of the calculated \EinA.

Since all lines measured in \textbf{Noll20} belong to the second and higher overtones, we might assume, based on our previous results \cite{Medvedev12,Medvedev16,Medvedev21,Medvedev22b,Ushakov23PN,Ushakov24}, that the problem is in  using non-analytic functions (\emph{e.g.} splines) for interpolation of the \ai\ data for the dipole-moment \cite{Stevens74,Mies74,Langhoff86,Turnbull89,Langhoff89,Goldman98,Loo08,Brooke16}\footnote{Analytical (polynomial \cite{Ferguson63,Nelson90}, Pad\'e approximant \cite{Chackerian92}, and exponential \cite{Medvedev95}) DMFs were also developed.}  
or potential-energy \cite{Stevens74,Brooke16} functions (DMFs or PEFs). However, earlier data of Franzen \emph{et al.} \cite{Franzen19} for lower $\Delta v=2$ and 3 also resulted in too low measured Q-branch \EinA\ as compared with HITRAN12.\footnote{For instance, the measured $A$ value of the 9-7Q$_1$(4) line is $4.7\pm0.2$ s$^{-1}$ \emph{vs} 8.3 s$^{-1}$ from HITRAN12.} This means that the problem is even more complicated.

Earlier, French \emph{et al.} \cite{French00} measured the Q/P and R/P airglow emission ratios 
to be lower than those calculated in Refs. \cite{Mies74,Langhoff86,Turnbull89}. Pendleton and Taylor \cite{Pendleton02} suggested that the primary source of this disparity is the coupling between the $X$ and $A$ states.

At this junction, it is in order to briefly describe three lowest electronic states of OH and the relevant perturbations. The $X^2\Pi_\Omega$ ground state is split by the spin-orbit interaction, $V_\textrm{so}$, into multiplet components, $F=1$ ($\Omega=3/2$) and $F=2$ ($\Omega=1/2$). Transitions $F_1\rightarrow F_1$ and $F_2\rightarrow F_2$ contribute to the main $X$-$X$ bands, the $F_1\leftrightarrow F_2$ transitions generate the fainter satellite bands \cite{Mies74,Brooke16}. Rotation mixes the multiplet components, therefore, two coupled Schr\"odinger equations have to be solved in order to obtain the vibrational functions of the $X$ state. The rotational perturbation in junction with the spin-orbit interaction mixes the $X^2\Pi_{1/2}$ with the lowest $A^2\Sigma^+_{1/2}$ excited state, thereby splitting the $X$ levels into the $\Lambda$ doublets and the $A$ levels into the $\Omega$ doublets. De Vivie \emph{et al.} \cite{deVivie88} considered also interactions with higher excited states and found that they essentially affected levels and their $\Lambda$ splittings at $v^\prime>5$ and high $J$.

Returning to the problem, we see that various approaches can be used to find the vibrational wave functions. Brooke \emph{et al.} \cite{Brooke16} solved a single Schr\"odinger equation, in which the effective Hamiltonian included all the above perturbations. 
The line strengths for the Meinel bands in HITRAN2020 \cite{Gordon22} are those calculated by \textbf{Brooke16} from two \emph{ab initio} dipole moment functions and the RKR potential curve. 
The most of the authors \cite{Mies74,Langhoff86,Nelson89a,Turnbull89,Chackerian92,Goldman98,Loo08} solve two coupled equations for the components of the $X$ multiplet. Mies \cite{Mies74} proposed the system of three coupled equations without explicitly solving them; he neglected the $A$-state contribution for $J<20$ and the energies below half of the dissociation limit. Mitev \emph{et al.} \cite{Mitev24} solved the system of three coupled equations to calculate the spectrum, yet no transition intensities were analyzed.

In this paper, we make the next step toward improving the intensities by solving the system of three coupled electronic states, $X^2\Pi_\textrm{3/2},X^2\Pi_\textrm{1/2}$, and $A^2\Sigma^+$, to calculate both the spectrum and the line intensities. The \ai\ calculations were performed for the BO potentials and other molecular functions in a wide range of the inter-nuclear distances. These data along with the experimental and theoretical literature data are used to fit the parameters of the model analytical PEFs and DMFs. The \EinA\ were calculated and analyzed. 

Section \ref{ham} describes the well-known Hamiltonian matrix for the $3\times3$ problem to introduce notations and to outline the features specific for the present work. In Sec. \ref{medmo}, the very cumbersome formulae for the matrix elements necessary to calculate the transition intensities are derived. Section \ref{modelmodel} presents the analytical forms to model various molecular functions with parameters to be fitted to the datasets described in Sec. \ref{data}, where our \ai\ calculations are also described. The fitting results in terms of the standard and reduced deviations are formulated in Sec. \ref{fitpro}. The applications of the theory to the problem of the variations in the level populations derived from the observational data are presented in Sec. \ref{PopDif}. The predicted intensity distributions of the $v$-0 low- and high-$J$ transitions as functions of $v$ and comparison with the \Brooke\ data are given in Sec. \ref{NIDLs}; the effect of the $X$-$A$ coupling is described explicitly. The conclusions in Sec. \ref{Concl} include description of the problems still remaining unresolved in the framework of the present model. In the appendix, we give a simple analytic explanation of the weakness of the satellite transitions. The Supplementary material contains the calculated \ai\ molecular functions and line list, the FORTRAN code to calculate the molecular functions with fitted parameters (including all necessary digits) and their plots as functions of $r$, and the fitting results.

\section{The Hamiltonian matrix for three coupled electronic states}\label{ham}

The general derivation of the Hamiltonian matrix in diatomic molecules is well described by Mies \cite{Mies74}. Here, we will emphasize 
some additional features used in our approach.

Let $(x,y,z)$ be the right-handed laboratory-fixed coordinate system (LCS) and $(\xi,\eta,\zeta)$ the right-handed molecule-fixed system (MCS) with origin $O$ at the nuclear center of gravity and the $O\zeta$ axis along the molecular axis whose orientation is defined by the polar angle, $\theta$, and azimuth, $\varphi$; $O\eta$ is in the $xy$ plane pointing in the direction of the vector product of the unit vectors along $O\zeta$ and $Oz$, and $O\xi$ is directed so as the MCS be right-handed; notation $q=(\xi,\eta,\zeta)$ is used for a set of all electronic coordinates in the MCS, and $r$ is the inter-nuclear distance.

The full set (truncated to three) of the adiabatic Born-Oppenheimer (BO) potentials and ortho-normalized wave functions as functions of $q$ and $r$ is found as solutions to the equation
\begin{equation}
    \left(T_{\textrm{el}} + V_{\textrm{Coul}} \right) \phi_n(q,r) = V_n^{\textrm{BO}}(r)\phi_n(q,r), \label{Tel}   
\end{equation}
where $n = X(\Lambda$=$1), A(\Lambda$=$0), X(\Lambda$=$-1$), and $\Lambda$ is eigenvalue of $L_\zeta$; we will use $n=n_1,n_2,n_3$, respectively. 
Note that $V_{n_1}^{\textrm{BO}}\equiv V_{n_3}^{\textrm{BO}}$ (to be denoted $V_{X}^{\textrm{BO}}$) since reflection in the molecular plane does not change the electronic Hamiltonian in Eq. (\ref{Tel}) but replaces $\Lambda$ with $-\Lambda$, the well-known $\Lambda$ degeneracy of the $\Pi$ state; $V_{n_2}^{\textrm{BO}}$ will be denoted $V_{A}^{\textrm{BO}}$. 
The model PEFs are to be fitted to the \ai\ points and to the experimental line positions. 

Let $\Pi_\xi$ and $\Pi_\eta$ stand for the real-valued normalized \ai\ wave functions, which transform as components of the polar vector. Let us write the normalized wave functions with $\Lambda=\pm1$ as
\begin{equation}
    \phi_{n_1}=\frac{1}{\sqrt{2}}\left(\Pi_\xi+i\Pi_\eta \right), \textrm{     } \phi_{n_3}=-\frac{1}{\sqrt{2}}\left(\Pi_\xi-i\Pi_\eta \right). \label{phiLam}
\end{equation}
Thus, the phases of functions $\phi_n$ are selected such that the following relations were true:
\begin{equation}
    \sigma_{\xi\zeta} \left|\phi_{n_1}\right> = -\left|\phi_{n_3}\right>, \sigma_{\eta\zeta} \left|\phi_{n_1}\right> = \left|\phi_{n_3}\right>, \label{sig_v}
\end{equation}
where $\sigma_{\xi\zeta}$ and $\sigma_{\eta\zeta}$ are reflections of the electronic coordinates in the respective molecular planes.

The adiabatic functions, $\phi_n$, depending only on the space coordinates must be combined with the spin functions, $\left|S,\Sigma\right>_\zeta$ ($S=\tfrac{1}{2},\Sigma=\pm\tfrac{1}{2}$; $\Sigma$ is projection onto the $\zeta$-axis), and with the rotational wave functions,
 $\left|JM\Omega\right> (\Omega=\Lambda+\Sigma)$, to give the full set of six non-symmetry-adapted (\emph{i.e.} without definite parity) electronic-rotational functions, the so-called Hund case (a) basis,
\begin{equation}
    \chi_i(\theta,\varphi,q;r) = \phi_n\left|S,\Sigma\right>_\zeta\left|JM\Omega\right>, \textrm{   } i=1,...,6,
    \label{nJM}
\end{equation} 
where the rotational wave functions \cite{Varshalovich88} depend on $\theta$ and $\varphi$. 
These functions with their specific parameters $n,\Lambda,\Sigma$, and $\Omega$  are presented in Table \ref{tabchi}; quantum numbers $J,M$ common for all functions are omitted.
\begin{table}[htbp]
    \centering
    \caption{Non-symmetry-adapted basis functions}
    \begin{tabular}{c|c|c|c|c|c|c}
         \hline
         & $\chi_1$ & $\chi_2$ & $\chi_3$ & $\chi_4$ & $\chi_5$ & $\chi_6$ \\
         \hline
        $n$ & $n_1$ & $n_1$ & $n_2$ & $n_2$ & $n_3$ & $n_3$ \\
    $\Lambda$     & 1 & 1 & 0 & 0 & -1 & -1 \\
    $\Sigma$      & +1/2 & -1/2 & +1/2 & -1/2 & +1/2 & -1/2 \\
    $\Omega$      & +3/2 & +1/2 & +1/2 & -1/2 & -1/2 & -3/2 \\
    \hline
    \end{tabular}
    \label{tabchi}
\end{table}

The total wave function can be expanded in this basis,
\begin{equation}
    \Psi(\theta,\varphi,q;r) = \frac{1}{r} \sum_i \chi_i(\theta,\varphi,q;r)\psi_i(r), \label{Psi}
\end{equation}
where the expansion coefficients, $\psi_i(r)$, are vibrational functions to be found from the Schr\"odinger equation,
\begin{equation}
    (H-E)\Psi = 0. \label{SchEq}
\end{equation}
In basis (\ref{nJM}), we obtain a system of six coupled equations for $\psi_i$. The $6\times6$ Hamiltonian matrix can be reduced to two $3\times3$ blocks by introducing the symmetry-adapted basis, \emph{i.e.} linear combinations of functions ($\ref{nJM}$) with definite parity, which will be performed later.

To proceed further, we need the matrix elements of the operators entering $H$ in the basis of functions (\ref{nJM}). First of all, we note that $H$ and all its components are scalars, therefore their matrix elements are diagonal in $M$ and are independent of $M$; in what follows, we will omit $M$ in the notations of the matrix elements of $H$.

Second, parameters $J,S,\Lambda,\Sigma,\Omega$ are ``good quantum numbers", \emph{i.e.} they have definite values for the Hund case (a) basis functions; in particular, $S=\tfrac{1}{2}$ for all of them.

In calculating the matrix elements of the rotational term
, we need to consider those of \textbf{J}, \textbf{L}, and \textbf{S}. Using the fact that components of vector \textbf{L} are Hermitian, we introduce the notation
\begin{equation}
    \Tilde{L}(r) \equiv \left<\phi_{n_1} \left|L_+\right|\phi_{n_2} \right> = \left<\phi_{n_2} \left|L_-\right|\phi_{n_1} \right>^*, \label{Ltilde0}
\end{equation}
where $L_\pm=L_\xi\pm iL_\eta$ and star stands for complex conjugate. It can be shown that function $\Tilde{L}(r)$ is real-valued as a direct consequence of the phase choice in Eqs. (\ref{sig_v}), it will be computed by the \ai\ methods \cite{Coxon75,Langhoff84,deVivie88}. Applying reflections (\ref{sig_v}) to Eq. (\ref{Ltilde0}), we obtain
\begin{IEEEeqnarray}{rcl}
    \left<\phi_{n_1} \left|L_+\right|\phi_{n_2} \right> &=& 
    \left<\phi_{n_2} \left|L_+\right|\phi_{n_3} \right> = \nonumber \\
    \left<\phi_{n_3} \left|L_-\right|\phi_{n_2} \right> &=&
    \left<\phi_{n_2} \left|L_-\right|\phi_{n_1} \right> = \Tilde{L}. \label{Ltilde}
\end{IEEEeqnarray}
Two more real-valued functions to be computed \ai\ are
\begin{eqnarray}
    \textbf{L}^2_X(r) &=& \left<\phi_n\right|\textbf{L}^2\left|\phi_n\right>,\hspace{10pt} n=n_1,n_3, \nonumber\\
    \textbf{L}^2_A(r) &=& \left<\phi_{n_2}\right|\textbf{L}^2\left|\phi_{n_2}\right>. \label{LAX}
\end{eqnarray}
    
The spin-orbit interaction $V_{\textrm{so}}$ can be characterized by two functions resulting from integrating $V_{\textrm{so}}$ with the electronic wave functions. 
The non-vanishing diagonal matrix elements of $V_{\textrm{so}}$ are 
\begin{IEEEeqnarray}{rcl}
    \left< \chi_1 \left|V_\textrm{so}\right| \chi_1 \right> &=& \left< \chi_6 \left|V_\textrm{so}\right| \chi_6 \right> = 
    + \tfrac{1}{2} A_{X}(r),  \nonumber  \\
    \left< \chi_2 \left|V_\textrm{so}\right| \chi_2 \right> &=& \left< \chi_5 \left|V_\textrm{so}\right| \chi_5 \right> = -\tfrac{1}{2} A_{X}(r). \label{Vsodiag}
\end{IEEEeqnarray}
where 
$A_X(r)$ is a real function of $r$ to be computed \ai\  \cite{Coxon75,Coxon82,Langhoff84,deVivie88,Borkov20}. 
The non-vanishing off-diagonal matrix elements 
are 
\begin{equation}
    \left< \chi_2 \left|V_\textrm{so}\right| \chi_3 \right> =  
    \left< \chi_4 \left|V_\textrm{so}\right| \chi_5 \right> = \tfrac{1}{2} A_{XA}(r), \label{Vsooff}
\end{equation}
where $A_{XA}(r)$ is also a real function of $r$ to be computed \ai\  \cite{Coxon75,Langhoff84,deVivie88}. 

The rotational term has the standard form, $B(r)\textbf{N}^2$, where $B=\hbar^2/2\mu r^2$, $\mu$ is the reduced nuclear mass and $\textbf{N}=\textbf{J}-\textbf{L}-\textbf{S}$ is the nuclear rotational angular momentum.

The diagonal matrix elements of $T_\textrm{vib}$ can be included in the effective BO potentials of the $X$ and $A$ states as the adiabatic mass-dependent corrections, therefore they are not considered here explicitly; the off-diagonal terms vanish between the electronic states of different symmetry \cite{Landau77}.

Further, Eq. (\ref{SchEq}) is converted to a $6\times6$ matrix form,
\begin{equation}
    \left[T_\textrm{vib} + \left( \begin{tabular}{cccccc}
       $U_1$ & $V$ & $Q$ & 0 & 0 & 0 \\
       $V$ & $U_2$ & $W$ & R & 0 & 0 \\
       $Q$ & $W$ & $U_3$ & $P$ & $R$ & 0 \\
       0 & R & $P$ & $U_3$ & $W$ & $Q$ \\
       0 & 0 & $R$ & $W$ & $U_2$ & $V$ \\
       0 & 0 & 0 & $Q$ & $V$ & $U_1$
    \end{tabular} \right) - E\right] \label{Hmatrix1} \left( \begin{tabular}{c}
         $\psi_1$  \\
         $\psi_2$  \\
         $\psi_3$  \\
         $\psi_4$  \\
         $\psi_5$  \\
         $\psi_6$ 
    \end{tabular} \right) = 0,
\end{equation}
where
\begin{IEEEeqnarray}{ccl}
    U_1 &=& \, V^\textrm{eff}_{X}+B\left(z_J^2-1\right)+\tfrac{1}{2}A_X,\textrm{  (term $X^2\Pi_{3/2}$)}, \label{U1}\\
    U_2 &=& \, V^\textrm{eff}_{X} + B\left(z_J^2+1\right)-\tfrac{1}{2}A_X,\textrm{  (term $X^2\Pi_{1/2}$)}, \label{U2}\\
    U_3 &=& \, V^\textrm{eff}_{A} + B\left(z_J^2+1\right),\hspace{33pt}\textrm{ (term $A^2\Sigma^+$}), \\
    V^\textrm{eff}_{X} &=& \, V^\textrm{BO}_{X}+B\left(\textbf{L}^2_X-1\right), \\
    V^\textrm{eff}_{A} &=& \, V^\textrm{BO}_{A}+B\textbf{L}^2_A, \\
    V &=& \, -Bz_J, \label{V}\\
    W &=& \,  B\Tilde{L}+\tfrac{1}{2}A_{XA}, \\
    R &=& \, -B\Tilde{L}\left(J+\tfrac{1}{2}\right), \\
    Q &=& \, -B\Tilde{L}z_J, \\
    P &=& \, -B\left(J+\tfrac{1}{2}\right),\\
    z_J &=& \, \sqrt{\left(J-\tfrac{1}{2}\right)\left(J+\tfrac{3}{2}\right)}. \label{U123}
\end{IEEEeqnarray}

Next, we introduce the symmetry-adapted basis by considering the transformation properties of the wave functions with respect to the operation of the \emph{total} inversion of the electronic and nuclear positions, \emph{I}, relative to $O$. Let us first perform the \emph{partial} inversion, $I_\textrm{n}$, which replaces the nuclear position vectors with their negatives while keeping the LCS unchanged. 
However, the MCS changes to MCS$^\prime$ due to rotation of nuclei around $O$. Namely, according to the definition given at the beginning of this section, the $O\zeta^\prime$ axis changes its direction to the opposite one with respect to $O\zeta$,
the $O\eta^\prime$ axis also changes its direction whereas the $O\xi$ axis \emph{remains unchanged}. 
Thus, MCS$^\prime$ is obtained from MCS by rotation MCS around $\xi$-axis by angle $\pi$. For definiteness, the rotation is performed counter-clockwise, which corresponds to increase of $\varphi$ by $+\pi$ under nuclear inversion.
As a result, the electronic coordinates ($q=+\xi,+\eta,+\zeta$) changed to $q=+\xi,-\eta,-\zeta$, the electrons themselves being still in their initial positions.\footnote{A notable difference between the particles' physical positions and their coordinates.} 

In order to complete the $I$ operation, one has to make the second \emph{partial} inversion, $I_\textrm{el}$, by replacing $q$ with their negatives, which results in $q=-\xi,+\eta,+\zeta$. Thus, the total inversion, $I=I_\textrm{el}I_\textrm{n}$, changes the electronic \emph{coordinates} $+\xi$ to $-\xi$, which is equivalent to reflection, $\sigma_{\eta\zeta}$, in the $\eta\zeta$ plane. 
Consider the action of \emph{I} on each factor of the total wave function separately.

The action of \emph{I} on the electronic $\phi_n$ functions is defined by Eqs. (\ref{sig_v}). 
Note that this operation changes $\Lambda$ to $-\Lambda$.

The transformation of the spin wave function is determined by the fact that spin is tightly bound to the molecule axis in Hund's case (a). This means that the spin projection onto the $O\zeta^\prime$ axis is the same as onto $O\zeta$, \emph{i.e.} $\Sigma^\prime=\Sigma$. In order to determine how the basis spin functions in the MCS$^\prime$, $\left|S,\Sigma^\prime\right>_{\zeta^\prime}$, are expressed in terms of the MCS functions, $\left|S,\Sigma\right>_\zeta$, we will write them as linear combinations of the functions with definite projections onto the $O\xi$ axis, $\left|S,\pm\frac{1}{2}\right>_\xi$,
\begin{eqnarray}
    \left|S,+\tfrac{1}{2}\right>_{\zeta^\prime} &=& \tfrac{1}{\sqrt{2}}\left(\left|S,\tfrac{1}{2}\right>_\xi + \left|S,-\tfrac{1}{2}\right>_\xi\right), \nonumber\\
    \left|S,-\tfrac{1}{2}\right>_{\zeta^\prime} &=& \tfrac{1}{\sqrt{2}}\left(\left|S,\tfrac{1}{2}\right>_\xi - \left|S,-\tfrac{1}{2}\right>_\xi\right)
    \label{spinbasis}
\end{eqnarray}
The respective original functions can be found by performing back rotation of MCS$^\prime$ to the MCS by angle $-\pi$ around $O\xi$. Under such rotation, functions $\left|S,\pm\frac{1}{2}\right>_\xi$ receive the phase factors $\exp{\left(\mp i\pi/2\right)}$. Then, it follows from Eqs. (\ref{spinbasis}) that
\begin{eqnarray}
    \left|S,+\tfrac{1}{2}\right>_{\zeta^\prime} &\rightarrow& e^{-i\pi/2}\left|S,-\tfrac{1}{2}\right>_\zeta, \nonumber \\
    \left|S,-\tfrac{1}{2}\right>_{\zeta^\prime} &\rightarrow& e^{-i\pi/2}\left|S,+\tfrac{1}{2}\right>_\zeta.
\end{eqnarray}
Note that inversion changes $\Sigma$ to $-\Sigma$.

The transformation of the rotational wave function is defined by the fact that rotation of the molecular axis under inversion \emph{I} is equivalent to the changes of the polar angle and azimuth, $\theta\rightarrow\pi-\theta$, $\varphi\rightarrow\varphi+\pi$. 
As follows from Eqs. 4.3(1) and 4.4(1) of Ref. \cite{Varshalovich88},
\begin{equation}
    I\Psi_{J,M,\Omega}(\theta,\varphi) \equiv \Psi_{J,M,\Omega}(\pi-\theta,\varphi+\pi) = e^{i\pi J} \Psi_{J,M,-\Omega}(\theta,\varphi).
    \label{psitot}
\end{equation}
Thus, under inversion, $\Omega$ changes its sign and, for the basis functions, the relation $\Omega=\Lambda+\Sigma$ is preserved.

Finally, summarizing the above findings, we obtain the following relations for the total wave functions:
\begin{equation}
    I\chi_1=(-1)^{J-1/2}\chi_6,\textrm{     }I\chi_2=(-1)^{J-1/2}\chi_5,\textrm{     }I\chi_3=(-1)^{J-1/2}\chi_4,
\end{equation}
and similarly for the other three.

Now, we can introduce the symmetry-adapted basis functions. Using notations \emph{e} and \emph{f} \cite{Kopp67,Brown75} for levels with parities $P=+\left(-1\right)^{J-1/2}$ and $-\left(-1\right)^{J-1/2}$, respectively,
we obtain
\begin{IEEEeqnarray}{rcl}
    \varphi_{1e} &=& 2^{-1/2}\left(\chi_1+\chi_6\right),\textrm{     }\varphi_{1f} = 2^{-1/2}\left(\chi_1-\chi_6\right), \nonumber\\
    \varphi_{2e} &=& 2^{-1/2}\left(\chi_2+\chi_5\right),\textrm{     }\varphi_{2f} = 2^{-1/2}\left(\chi_2-\chi_5\right), \nonumber\\
    \varphi_{3e} &=& 2^{-1/2}\left(\chi_3+\chi_4\right),\textrm{     }\varphi_{3f} = 2^{-1/2}\left(\chi_3-\chi_4\right). \label{sabasis}
\end{IEEEeqnarray}
Transforming the matrix elements of $H$, Eq. (\ref{Hmatrix1}), to this basis, we obtain
\begin{equation}
    \left[T_\textrm{vib} + \left( \begin{tabular}{cccccc}
       $U_1$ & $V$ & $Q$ & 0 & 0 & 0 \\
       $V$ & $U_2$ & $W+R$ & 0 & 0 & 0 \\
       $Q$ & $W+R$ & $U_3+P$ & 0 & 0 & 0 \\
       0 & 0 & 0 & $U_1$ & $V$ & $Q$ \\
       0 & 0 & 0 & $V$ & $U_2$ & $W-R$ \\
       0 & 0 & 0 & $Q$ & $W-R$ & $U_3-P$
    \end{tabular} \right) - E\right] \label{Hmatrix2} \left( \begin{tabular}{c}
         $\psi_{1e}$  \\
         $\psi_{2e}$  \\
         $\psi_{3e}$  \\
         $\psi_{1f}$  \\
         $\psi_{2f}$  \\
         $\psi_{3f}$ \\         
    \end{tabular} \right) = 0,
\end{equation}
where $\psi_{1e}$,...,$\psi_{3f}$ are coefficients of expansion of function $\Psi(\theta,\varphi,q;r)$ in the symmetry-adapted basis (\ref{sabasis}), 
\begin{equation}
    \Psi(\theta,\varphi,q;r) \equiv \Psi_{S,J,M,p} = \frac{1}{r}\sum_{i=1}^3 \varphi_{ip}\psi_{ip}, \hspace{20pt} p=e,f. \label{Psief}
\end{equation}
The radial amplitudes in Eq. (\ref{Hmatrix1}), $\psi _{i}$, are related to the symmetrized
functions, $\psi _{kp}$, by the relations 
\begin{IEEEeqnarray}{rcl}
    \psi _{1}=2^{-1/2}(\psi _{1e}+\psi _{1f}),\quad \psi _{6}=2^{-1/2}(\psi _{1e}-\psi _{1f}), \nonumber\\
    \psi _{2}=2^{-1/2}(\psi _{2e}+\psi _{2f}),\quad \psi _{5}=2^{-1/2}(\psi _{2e}-\psi _{2f}), \nonumber\\
    \psi _{3}=2^{-1/2}(\psi _{3e}+\psi _{3f}),\quad \psi _{4}=2^{-1/2}(\psi _{3e}-\psi _{3f}). \label{sabasisR}
\end{IEEEeqnarray}

As a result, the Hamiltonian matrix is decomposed into two 3$\times$3 blocks of the $e$ and $f$ types with opposite parities.
Thus, depending on the parity, there are three components of the total wave functions (\ref{Psief}) of the $e$ type or three of the $f$ type to be used in calculations of the transition matrix elements in the next section.

It should be noted that six equations (\ref{Hmatrix2}) are to be solved at each $J$ except for $J=\tfrac{1}{2}$, in which case states with $\Omega=\tfrac{3}{2}$ do not exist and only four equations for $\psi_{2e},\psi_{3e},\psi_{2f},\psi_{3f}$ remain.

\section{The matrix elements of the dipole-moment operator} \label{medmo}

The transition dipole moment (TDM) of the optical transition between the upper state (primed) and lower state (double-primed) is defined by the $z$ component of the dipole-moment operator, $\textbf{d}(\theta,\varphi,q;r)$, in the LCS, 
\begin{eqnarray}
\text{TDM} &=& \int \Psi ^{\prime } d_z \Psi ^{\prime
\prime
}\, \,d\cos \theta \, d\varphi\,dq\,r^2dr  \notag \\
&=& \int \sum_{i,k=1}^{6}\psi _{i}^{\prime }(r)\chi _{i}(\theta
,\varphi ,q;r)\,d_z\,\psi _{k}^{\prime \prime }(r)\chi
_{k}(\theta ,\varphi
,q;r) \,d\cos \theta \, d\varphi\,dq\,dr  \notag \\
&=&\int \sum_{i,k=1}^{6}\psi _{i}^{\prime }(r)\,\left\langle \chi
_{i}\left\vert d_{z}\right\vert \chi _{k}\right\rangle \,\psi
_{k}^{\prime \prime }(r)\, dr\,.  \label{TDMgeneral}
\end{eqnarray}%
where $\left\langle \chi _{i}\left\vert d_{z}\right\vert \chi
_{k}\right\rangle$ are the matrix elements of $d_z$
in the basis of the non-symmetry-adapted electronic-rotational functions specified in Eq.
(\ref{nJM}) and Table \ref{tabchi}.

The above matrix elements will contain contributions from all three components of \textbf{d} in the MCS, that is, $d_\xi(q,r),d_\eta(q,r),d_\zeta(q,r)$. The LCS $z$ component is expressed \emph{via} the latters as
\begin{equation}
    d_z = \tfrac{1}{2}\left(d_++d_-\right) \sin{\theta} + d_\zeta\cos{\theta},    \label{dz}
\end{equation}
where $d_\pm=d_\xi\pm id_\eta$.

In calculations of the TDMs, we have first to 
calculate the matrix elements of $d_{\pm}$ and $d_{\zeta}$ in the basis of the MCS BO electronic functions, $\phi_{n}$, which gives three DMFs as functions of $r$. 
We introduce the following notations for two DMFs diagonal in the \emph{X} and $A$ states,
\begin{IEEEeqnarray}{ccl}
    d^X(r) &=& \left<\phi_{n_1}\left|d_\zeta\right| \phi_{n_1} \right> = \left<\phi_{n_3}\left|d_\zeta\right| \phi_{n_3} \right>, \nonumber\\
    d^A(r) &=& \left<\phi_{n_2}\left|d_\zeta\right| \phi_{n_2} \right>,
        \label{DMFdiag}
\end{IEEEeqnarray}
and one off-diagonal DMF,
\begin{eqnarray}
    d^{XA}(r) &=& \left<\phi_{n_1}\left|d_+\right| \phi_{n_2} \right> = \left<\phi_{n_2}\left|d_-\right| \phi_{n_1} \right>, \nonumber\\
    &=& -\left<\phi_{n_2}\left|d_+\right| \phi_{n_3} \right> = -\left<\phi_{n_3}\left|d_-\right| \phi_{n_2} \right>.
    \label{DMFoffdiag}
\end{eqnarray}
All three DMFs are real-valued and are to be computed \ai. The relative signs and the real-valuedness of the matrix elements in Eqs. (\ref{DMFdiag}) and (\ref{DMFoffdiag}) are ensured by the phase conventions in Eqs. (\ref{sig_v}).

The selection rules and the values of matrix elements $\left\langle
\chi _{i}\left\vert d_{z}\right\vert \chi _{k}\right\rangle$ are
obtained with standard application of the Wigner-Eckart theorem.
Using the notation
\begin{equation}
\left\langle \chi _{i}\left\vert d_{z}\right\vert \chi
_{k}\right\rangle \equiv \left\langle J^{\prime },M^{\prime },\Omega
^{\prime },\Lambda ^{\prime }\left\vert d_{z}\right\vert
J^{\prime\prime},M^{\prime\prime},\Omega ^{\prime\prime},\Lambda ^{\prime\prime}\right\rangle \equiv \left(
d_{z}\right) _{J^{\prime\prime},M^{\prime\prime},\Omega ^{\prime\prime},\Lambda ^{\prime\prime}}^{J^{\prime
},M^{\prime },\Omega ^{\prime },\Lambda ^{\prime }}\,, \label{medz}
\end{equation}%
we write down the non-zero matrix elements (\ref{medz}) as products of the $M$-dependent factors and the reduced matrix elements as functions of $r$,
\begin{equation}
\left( d_{z}\right) _{J,M,\Omega ,\Lambda }^{J,M,\Omega ,\Lambda }=\frac{%
M\Omega }{J\left( J+1\right) }\,d^{B}(r),\hspace{5pt}B=X \textrm{  for  } \Lambda=\pm1, A \textrm{  for 
 } \Lambda=0,  \label{dIni}
\end{equation}%
\begin{eqnarray}
\left( d_{z}\right) _{J,M,\Omega -1,\Lambda -1}^{J,M,\Omega
,\Lambda } &=&
\left( d_{z}\right) _{J,M,\Omega ,\Lambda }^{J,M,\Omega -1,\Lambda -1}%
  \notag \\
&=&-(-1)^{\Lambda }\frac{M\sqrt{\left( J+\Omega \right) \left(
J-\Omega +1\right) }}{2J\left( J+1\right) }\,d^{XA}(r),
\end{eqnarray}%

\begin{eqnarray}
\left( d_{z}\right) _{J,M,\Omega ,\Lambda }^{J-1,M,\Omega ,\Lambda
} &=& \left[ \left( d_{z}\right) _{J-1,M,\Omega ,\Lambda }^{J,M,\Omega
,\Lambda }\right]
^{\ast }  \notag \\
&=&i\frac{{1}}{J}\sqrt{\frac{\left( J^{2}-M^{2}\right) \left(
J^{2}-\Omega ^{2}\right) }{4J^{2}-1}}\,d^{B}(r),\hspace{5pt}B\textrm{ as above,}
\end{eqnarray}%

\begin{eqnarray}
\left( d_{z}\right) _{J,M,\Omega -1,\Lambda -1}^{J-1,M,\Omega
,\Lambda } &=& \left[ \left( d_{z}\right) _{J-1,M,\Omega ,\Lambda
}^{J,M,\Omega -1,\Lambda
-1}\right] ^{\ast } \notag \\
&=&-i(-1)^{\Lambda }\frac{{1}}{2J}\sqrt{\frac{\left(
J^{2}-M^{2}\right) \left( J-\Omega \right) \left( J-\Omega +1\right)
}{4J^{2}-1}}\,d^{XA}(r), \notag\\
\end{eqnarray}%
\begin{eqnarray}
\left( d_{z}\right) _{J,M,\Omega ,\Lambda }^{J-1,M,\Omega
-1,\Lambda -1} &=& \left[ \left( d_{z}\right) _{J-1,M,\Omega -1,\Lambda
-1}^{J,M,\Omega
,\Lambda }\right] ^{\ast } \notag \\
&=&i(-1)^{\Lambda }\frac{{1}}{2J}\sqrt{\frac{\left(
J^{2}-M^{2}\right) \left( J+\Omega \right) \left( J+\Omega -1\right)
}{4J^{2}-1}}\,d^{XA}(r). \label{dFin} \notag\\
\end{eqnarray}%

Finally, we insert matrix elements (\ref{dIni})-(\ref{dFin}) and functions (\ref{sabasisR}) into the last line of Eq. (\ref{TDMgeneral}).
Then, the integrals arise of the three DMFs defined in Eqs. (\ref{DMFdiag}) and (\ref{DMFoffdiag}) with the vibrational wave functions, $\psi_{i^\prime p^\prime}^\prime$ and $\psi_{i^{\prime\prime}p^{\prime\prime}}^{\prime\prime}$,

\begin{IEEEeqnarray}{ccl}
    d^X_{i^\prime p^\prime,i^{\prime\prime}p^{\prime\prime}} &=& \int_0^\infty\psi^\prime_{i^\prime p^\prime}d^X\psi^{\prime\prime}_{i^{\prime\prime}p^{\prime\prime}}dr, \nonumber\\
    d^A_{i^\prime p^\prime,i^{\prime\prime}p^{\prime\prime}} &=& \int_0^\infty\psi^\prime_{i^\prime p^\prime}d^A\psi^{\prime\prime}_{i^{\prime\prime}p^{\prime\prime}}dr, \nonumber\\
    d^{XA}_{i^\prime p^\prime,i^{\prime\prime}p^{\prime\prime}} &=& \int_0^\infty\psi^\prime_{i^\prime p^\prime}d^{XA}\psi^{\prime\prime}_{i^{\prime\prime}p^{\prime\prime}}dr,\hspace{10pt}i^\prime,i^{\prime\prime}=1,2,3, \hspace{10pt}p^\prime,p^{\prime\prime} =e,f.
    \label{dXdAdXA}
\end{IEEEeqnarray}

Using the above expressions,
we find the TDMs for the P, Q, and R branches as follows. %

\vspace{10pt} \noindent \underline{P branch:
$J^\prime=J-1,J^{\prime\prime}=J$}
\begin{equation}
\text{TDM}=i\sqrt{\frac{J^{2}-M^{2}}{J\left( 2J+1\right) }}d \label{TDMPbranch}
\end{equation}
\underline{\emph{$e\rightarrow e$}}
\begin{eqnarray}
d &=&\frac{1}{4}\Big[ 2d_{1e,1e}^{X}\sqrt{4J^{2}-9}+\left(
2d_{2e,2e}^{X}+2d_{3e,3e}^{A}-d_{3e,2e}^{XA}+d_{2e,3e}^{XA}\right)\sqrt{4J^{2}-1} + 
\notag \\
&& d_{1e,3e}^{XA}\sqrt{4( J-1) ^{2}-1}-d_{3e,1e}^{XA}\sqrt{%
4( J+1) ^{2}-1}\Big] \Big/ \sqrt{J(2J-1)} \label{TDMPee}
\end{eqnarray}%
\underline{\emph{$f\rightarrow f$}}
\begin{eqnarray}
d &=&\frac{1}{4}\Big[ 2d_{1f,1f}^{X}\sqrt{4J^{2}-9}+\left(
2d_{2f,2f}^{X}+2d_{3f,3f}^{A}+d_{3f,2f}^{XA}-d_{2f,3f}^{XA}\right)\sqrt{4J^{2}-1} + 
\notag \\
&& d_{1f,3f}^{XA}\sqrt{4( J-1) ^{2}-1}-d_{3f,1f}^{XA}\sqrt{%
4( J+1) ^{2}-1}\Big] \Big/ \sqrt{J(2J-1)}  \label{TDMPeeff}
\end{eqnarray}%
\vspace{10pt} \noindent \underline{Q branch:
$J^\prime=J,J^{\prime\prime}=J$}
\begin{equation}
\textrm{TDM}=\frac{M}{\sqrt{J\left( J+1\right) }}\,d\, \label{TDMQbranch}
\end{equation}%
\underline{\emph{$e\rightarrow f$}}
\begin{eqnarray}
d &=&\frac{1}{4}\Big[ \left( {d}%
_{1e,3f}^{XA}+{d}_{3e,1f}^{XA}\right)\sqrt{\left( 2J+3\right) \left( 2J-1\right) } -\left( {d}%
_{2e,3f}^{XA}-{d}_{3e,2f}^{XA}\right)\left( 2J+1\right) +   \notag \\
&& 6d_{1e,1f}^{X}+2d_{2e,2f}^{X}+2d_{3e,3f}^{A}  \Big] \Big/ \sqrt{J(J+1)} \label{TDMQfeef}
\end{eqnarray}%
\underline{\emph{$f\rightarrow e$}}
\begin{eqnarray}
d &=&\frac{1}{4}\Big[ \left( {d}%
_{1f,3e}^{XA}+{d}_{3f,1e}^{XA}\right)\sqrt{\left( 2J+3\right) \left( 2J-1\right) } +\left( {d}%
_{2f,3e}^{XA}-{d}_{3f,2e}^{XA}\right)\left( 2J+1\right) +   \notag \\
&& 6d_{1f,1e}^{X}+2d_{2f,2e}^{X}+2d_{3f,3e}^{A}\Big] \Big/ \sqrt{
J(J+1)}  \label{TDMQeffe}
\end{eqnarray}%
\vspace{10pt} \noindent \underline{R branch:
$J^\prime=J,J^{\prime\prime}=J-1$}
\begin{equation}
\text{TDM}=-i\sqrt{\frac{J^{2}-M^{2}}{J\left( 2J-1\right) }}\,d\, \label{TDMRbranch}
\end{equation}
\underline{\emph{$e\rightarrow e$}}
\begin{eqnarray}
d &=&\frac{1}{4}\Big[ 2d_{1e,1e}^{X}\sqrt{4J^{2}-9}+\left(
2d_{2e,2e}^{X}+2d_{3e,3e}^{A}+d_{3e,2e}^{XA}-d_{2e,3e}^{XA}\right)\sqrt{4J^{2}-1} + 
\notag \\
&& d_{3e,1e}^{XA}\sqrt{4( J-1) ^{2}-1}-d_{1e,3e}^{XA}\sqrt{%
4( J+1) ^{2}-1}\Big] \Big/ \sqrt{J(2J+1)} \label{TDMRee}
\end{eqnarray}%
\underline{\emph{$f\rightarrow f$}}
\begin{eqnarray}
d &=&\frac{1}{4}\Big[ 2d_{1f,1f}^{X}\sqrt{4J^{2}-9}+\left(
2d_{2f,2f}^{X}+2d_{3f,3f}^{A}-d_{3f,2f}^{XA}+d_{2f,3f}^{XA}\right)\sqrt{4J^{2}-1} + 
\notag \\
&& d_{3f,1f}^{XA}\sqrt{4( J-1) ^{2}-1}-d_{1f,3f}^{XA}\sqrt{%
4( J+1) ^{2}-1} \Big] \Big/ \sqrt{J(2J+1)}  \label{TDMReeff}
\end{eqnarray}%
The $1\leftrightarrow2$ matrix elements of the DMFs between the $X$ states with $\Omega=3/2$ ($1e,1f$) and 1/2 ($2e,2f)$  are absent in the TDMs of Eqs. (\ref{TDMPbranch})-(\ref{TDMReeff}) because the dipole moment is independent of spin and $\Delta\Lambda=2$ is not allowed by the selection rules for the dipole transitions. This also explains the weakness of the satellite lines at relatively low $J$. When $J$ increases, integrals $d_{1p^\prime,1p^{\prime\prime}}$ and $d_{2p^\prime,2p^{\prime\prime}}$ become comparable in both the main and satellite transitions, which nevertheless remain weak due to their mutual cancellation, see Appendix.

Squares of these TDMs, after averaging over $M$ in the upper state,
are equal to $d^2/3$. As a result, for the \EinA, we get for all
branches
\begin{equation}
A=3.1361891\times 10^{-7}\nu ^{3}\,d^{2}, \label{EinA}
\end{equation}
where $A$ is in s$^{-1}$, $\nu$ in cm$^{-1}$, and $d$ in
debye.

It is seen that the non-vanishing matrix elements in the above equations are present for the $e\leftrightarrow e$ and $f\leftrightarrow
f$ transitions in the P and R branches, the $e\leftrightarrow f$ and
$f\leftrightarrow e$ transitions in the Q branch, which is a consequence of the fact that
operator $d_z$ has negative parity.

\section{The model functions} \label{modelmodel}

Following our general strategy \cite{Meshkov18,Meshkov22,Ushakov23PN,Medvedev24MP}, we use analytical functions for modelling various functions of $r$ entering the Hamiltonians for the elecronic $X$ and $A$ states. We remind that the analytical functions have continuous derivatives of all orders and, as a consequence, possess valuable properties, of which the most important one in calculations of the overtone-transition intensities is the possibility of analytical continuation of the functions from the real axis, where all of them are usually specified initially, into the complex plane. The intensities of such transitions are exponentially decreasing with the overtone number \cite{Medvedev12} whereas the integrand in the TDM integral is not, \emph{i.e.} severe cancellation occurs that results in the integral values orders of magnitude smaller than the maximum values of the integrand. As shown in the textbook by Landau and Lifshitz \cite[\S87]{Landau77}, in such a case, 
the exponentially small TDM values are due to contributions of the integrand singularities, \emph{i.e.} poles and branch points, in the complex plane.
The important practical recommendation is that the singularity providing for the largest contribution be not artificial, depending on the model choice; rather, it must have physical meaning, as do, \emph{e.g.}, the strong repulsion at small $r$ in the PEF. In contrast, the functions with discontinuities in derivatives may result in a non-physical behavior of the overtone intensities as demonstrated in Ref. \cite{Medvedev22b}.

For modelling the BO potentials, $V^{\textrm{BO}}_X(r)$ and $V^{\textrm{BO}}_A(r)$, we employed Meshkov's function \cite{Meshkov18} with different mapping functions. Omitting the index of the electronic state, we write the model PEFs in the form
\begin{IEEEeqnarray}{rcl}
    V^{\textrm{BO}}(r) &=& U_0+\left\{\left[\frac{K}{r}+E_0+K\alpha+ \left(E_0+\tfrac{1}{2}K\alpha\right) \alpha r\right]e^{-\alpha r}\right. \nonumber\\
                   &+&\left. \left[1-\exp{\left(-d_1r-d_2r^2\right)} \right]^8 \frac{C_6} {r^6}\right\}\left(1+x^3\sum_{i=0}^{12} a_ix^i \right), \\
                  x &=& \tanh{(\alpha r)},\hspace{7pt}\sum_{i=0}^{12}a_i=-2. \nonumber
                   \label{VBOn}
\end{IEEEeqnarray}
The potential contains 21 parameters, including 17 that are adjusted in the fitting. 
Constant $U_0$, \emph{i.e.} the asymptotic limit at $r\rightarrow\infty$, was put to zero for the $X$ state and $15867.86$ cm$^{-1}$, the excitation energy of the O($^1$D) atom, in the $A$ state \cite{Radzig85,Mitev24}.
Constants $K$ and $E_0$ were fixed in accord with the united-atom limits in the $X$ and $A$ states, whereas two sets of $\alpha,C_6,d_1,d_2,a_i$ for $X$ and $A$ were variable. 

In the Hamiltonian matrix, correction terms were added to the BO potentials in order to describe the observed energy levels. 
The necessity of such corrections stems in the fact established by the calculations of De Vivie \emph{et. al.} \cite{deVivie88} that the vibrational levels of the $A$ state with $v>5$ are essentially affected by intersections of the $A$ term with three repulsive electronic terms converging to the same asymptotic limit at $r\rightarrow\infty$ as the $X$ term, which are not included in the present model.

Three types of the correction functions were used, 
\begin{equation}
    V_{\textrm{corr},a,n}(r) = \sum_{i=0}^nc_{a,i}\left[\tanh{(qr)}\right]^i, \label{Vcorra}
\end{equation}
\begin{equation}
    V_{\textrm{corr},b}(r) = \sum_{i=1}^8c_{b,i} 2^{-i}\left[1+\tanh{\left(q_1r-q_2r^{-1}\right)}\right]^i, \label{Vcorrb}
\end{equation}
\begin{equation}
    V_{\textrm{corr},c}(r) = \left(C_1y+C_2y^2\right)\left(\frac{r_0}{r}\right)^6, \hspace{5pt}y=1-\tanh{\left(\frac{r_0}{r}\right)^2}. \label{Vcorrc}
\end{equation}
The correction functions to the $X$ and $A$ potentials had the forms
\begin{eqnarray}
    V_{\textrm{corr},X}(r) &=& V_{\textrm{corr},a,8}(r), \label{VcorrX} \\
    V_{\textrm{corr},A}(r) &=& V_{\textrm{corr},a,10}(r)+V_{\textrm{corr},b}(r)+V_{\textrm{corr},c}(r). \label{VcorrA} 
\end{eqnarray}

The expectation values of $\textbf{L}^2$ in states $X$ and $A$ are introduced in Eqs. (\ref{LAX}). Omitting the electronic index, we write the model functions as
\begin{eqnarray}
   \textbf{L}^2(r) &=& \left(2+br^2\right)\left(1+x^2\sum_{i=0}^nl_ix^i \right), \\
     && x=\tanh{(lr)},\hspace{20pt}\sum_{i=0}^nl_i=0. \label{Lr} \nonumber
\end{eqnarray}
Parameter $n$ was set to 3 in the $X$ state and 5 in $A$. 

The off-diagonal matrix element of the electronic orbital angular momentum, Eqs. (\ref{Ltilde0}) and (\ref{Ltilde}), was modelled with the function
\begin{equation}
    \Tilde{L}(r) = \exp{\left(-\lambda_1r-\lambda_2r^2\right)} \sum_{i=0}^6 \Tilde{l}_ix^i, \hspace{17pt}x=\tanh{(\Tilde{l}r)}. \label{Ltilder}
\end{equation}

The diagonal and off-diagonal spin-orbit couplings, $A_X$ and $A_{XA}$, defined in Eqs. (\ref{Vsodiag}) and (\ref{Vsooff}) are both modelled with the function 
\begin{eqnarray}
    A(r) &=& 2\left(S_0+S_1e^{-\sigma r}\right)\left(1+\sum_{i=1}^7s_ix^i \right), \label{AXA} \\ 
     && x=\tanh{(sr)},\hspace{20pt}\sum_{i=1}^7s_i=0. \nonumber
\end{eqnarray}

All three DMFs were modelled with one and the same irregular function \cite{Medvedev22a} having the same number of parameters with individual values for each function,
\begin{equation}
d_\textrm{irreg}(r)=
\frac{\left(1-e^{-c_2r}\right)^3}{\sqrt{\left(r^2-c_3^2\right)^2+c_4^2}
\sqrt{\left(r^2-c_5^2\right)^2+c_6^2}}
\sum_{k=0}^{6}b_k \left(1-2e^{-c_1r}\right)^k. \label{sing2}
\end{equation}

Supplementary material contains the FORTRAN code for calculating all molecular functions with fitted parameters.

\section{The datasets used in the fitting}\label{data}

The fitting of the molecular functions specified in Sec. \ref{modelmodel} was performed to the following experimental and theoretical data.

1. The experimental energy levels in the $X$ and $A$ states \cite{Furtenbacher22} and the $X$-$X$ transition frequencies collected in \textbf{Brooke16} from various sources 
\cite{Maillard76,Sappey90,Copeland93,Abrams94,Melen95,Nizkorodov01,Bernath09,Martin11} (see Table 1 in Ref. \cite{Bernath09}).

2. The relative intensities of the transitions with a common upper level. Laboratory measurements of $\Delta v=1,2$ emission and absorption \cite{Nelson90,Nelson89b}. 

3. The relative intensities of the transitions with a common upper level. Astrophysical observations of the $\Delta v =3$-6 transitions in emission \cite{Noll20}.
 
4. The line intensities in the $\Delta v=2$ band recalculated in \Brooke\ from the spectra recorded by Abrams \emph{et al.} \cite{Abrams94}.

5. The permanent dipole moment, $\mu$, in the $v=0,$1,2 states \cite{Peterson84}. In fact, the TDMs for Q-branch transitions denoted $\mu_{ij}$ were measured and then converted to $\mu$. Using the data of \cite[Table 3]{Peterson84}, we found the original measured values of $\mu_{ij}$ and compared them with our calculated TDMs.

6. The \emph{ab initio} calculations of the present study (PS).
They were performed with MOLPRO-2010.1 program package \cite{Werner10}. All the model functions specified in Sec. \ref{modelmodel}
were calculated by Multi Reference Configuration Interaction (MRDCI) approach. The active space of MOs were prepared by CASSCF method with the state-average technique for one double occupied core MO and seven electrons distributed over nine active MOs. The aug-cc-pV5Z basis set was used at oxygen and hydrogen atoms. The functions were calculated  at $r=0.4$-8.0 bohr. The \ai\ data of the present study and the experimental data mentioned above are given in Supplementary material.

We don't use the recently measured absolute intensities in the fundamental band \cite{Chang23} since the experimental conditions are not specified in full detail.

\section{Fitting} \label{fitpro}

\begin{figure}[htbp]
    \centering
    \includegraphics[scale=0.4]{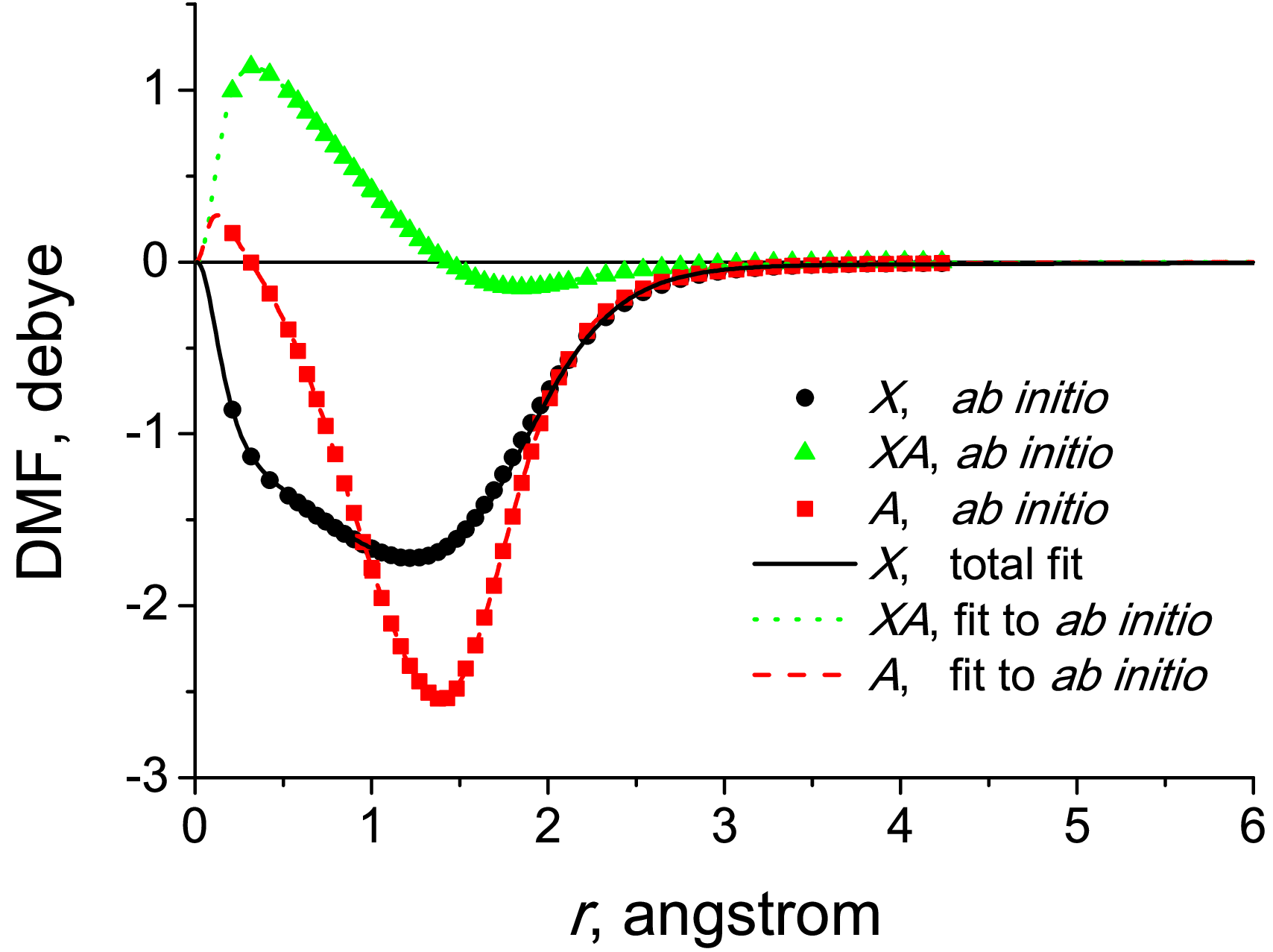}
    \caption{The \emph{ab initio} (points) and model (lines) dipole-moment functions. The model DMFs were fitted either to both the \emph{ab initio} and experimental data (total fit) or to \emph{ab initio} only (fit to \emph{ab initio}).}
    \label{figDMF}
\end{figure}

\begin{figure}[htbp]
    \centering
    \includegraphics[scale=0.4]{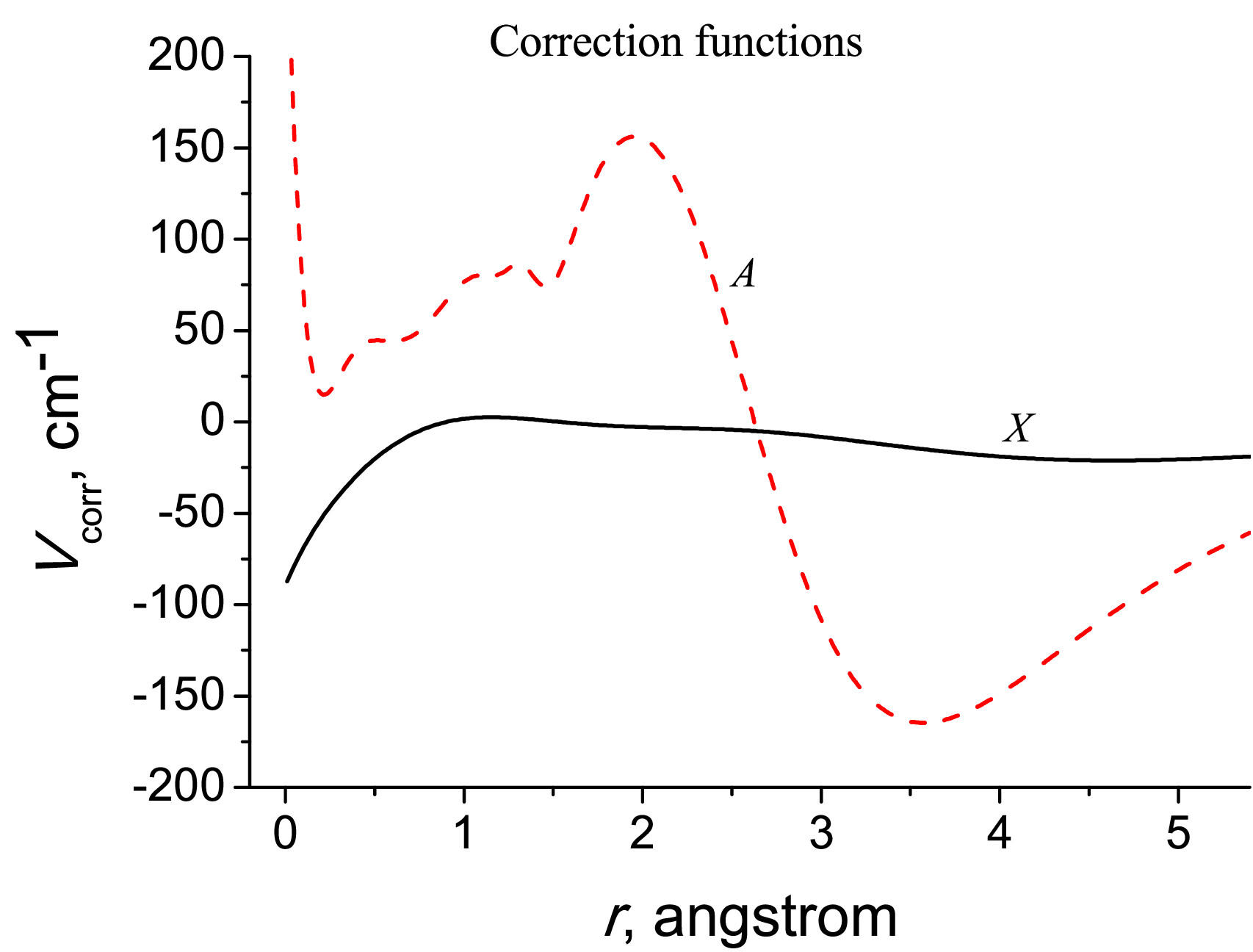}
    \caption{The corrections to the PEFs of the $X$ and $A$ states. The non-monotonic behavior of the $V_{\textrm{corr},A}$ at $r<2$ \AA\ is a shortcoming of the model.}
    \label{figVcorr}
\end{figure}

The fitting was performed in two steps. First, the BO potentials and other functions of Eqs. (\ref{VBOn})-(\ref{AXA}) were fitted to the experimental energy levels and transition frequencies; the total of about 110 parameters were adjusted.
Some of the fitted functions (lines) along with the \ai\ data (points) are shown in Figs. \ref{figDMF} and \ref{figVcorr}. 

Figure \ref{figDMF} demonstrates three fitted DMFs defined in Eqs. (\ref{DMFdiag})-(\ref{DMFoffdiag}) and modelled with the irregular function of Eq. (\ref{sing2}), which showed good results for CO \cite{Medvedev22a}. The $d^X$ model was fitted to both \ai\ data and to all the intensity data and permanent dipole, as indicated in points 2-6 of Sec. \ref{data}. The other two functions were fitted only to the \ai\ data because the total fit did not have appreciable effect. All three functions are smooth and have correct behavior in both short- and long-range regions.

Figure \ref{figVcorr} shows the correction functions that were added to the BO potentials in order to describe the observed line positions. The non-monotonic behavior of $V_{\textrm{corr},A}$ at $r<2$ \AA\ certifies a drawback of the three-states model. The same is true of $\Tilde{L}$ at $r<1.3$ \AA, see Fig. S1 in Supplementary material where other fitted molecular functions defined in Sec. \ref{modelmodel} are also shown.

The fitting results for the energy levels are shown in Table \ref{tab_rmsXA} along with the data of Mitev \emph{et al.} \cite{Mitev24}. For transition frequencies, the present study gives rmsd = 0.10 cm$^{-1}$. Both sets reproduce the most of the levels within 1 cm$^{-1}$, yet far beyond the experimental errors. In contrast, the theory of \Brooke\ gives the levels within the experimental uncertainties ($\chi_\textrm{red}=0.96$,\footnote{For definition, see Eq. (\ref{chi}).} rmsd = 0.026 cm$^{-1}$), therefore their line list provides for the best transition frequencies included in the list of the present study.
\begin{table}[htbp]
    \centering
    \caption{Root-meam-square deviations of the energy levels (cm$^{-1}$) \cite{Furtenbacher22} in the $X$ and $A$ states calculated in the present and previous studies}
    \vspace{10pt}
    \begin{tabular}{c|c|c|c|c}
        & \multicolumn{2}{c|}{Present study} & \multicolumn{2}{c}{Mitev \emph{et al.} \cite{Mitev24}}\\
       \hline 
     $v$ & rmsd$_X$ & rmsd$_A$ & rmsd$_X$ & rmsd$_A$ \\
     \hline
    0 & 0.23 & 0.26 & 0.08 & 0.07 \\
    1 & 0.24 & 0.39 & 0.09 & 0.19 \\
    2 & 0.18 & 1.02 & 0.15 & 1.05 \\
    3 & 0.12 & 0.87 & 0.08 & 0.76 \\
    4 & 0.07 & 0.97 & 0.12 & 0.61 \\
    5 & 0.06 & 0.69 & 0.06 & 1.96 \\
    6 & 0.06 & 0.59 & 0.11 & 3.23 \\
    7 & 0.06 & 0.78 & 0.14 & 4.70 \\
    8 & 0.08 & 6.48 & 0.08 & 1.64 \\
    9 & 0.13 &14.0  & 0.25 & 4.58 \\
   10 & 0.08 &      & 0.25 &      \\
   11 & 0.19 &      & 0.70 &      \\
   12 & 0.26 &      & 1.34 &      \\
   13 & 0.31 &      & 1.13 &      \\
   \hline
   Whole state & 0.17 & 3.49 & 0.33 & 1.79 \\
   \hline
    \end{tabular}
    \label{tab_rmsXA}
\end{table}
\begin{table}[htbp]
    \centering
    \caption{The fitting results for the intensities and permanent-dipole data from various source data used in the fitting}
    \vspace{12pt}
    \begin{tabular}{l|c|c|c}
      Parameter   &  $\chi_\textrm{red}$ & rms & source\\
      \hline
       Intensity ratios$^a$  & 0.59 & 0.008 & \Brooke \\
       Intensity ratios$^a$  & 0.71 & 3.46 & \Noll \\
       Intensity ratios$^b$  & 2.30 & 0.21 & \Noll \\
       Intensity ratios$^c$  & 0.90 & 1.26 & \cite{Nelson89b,Nelson90} \\
       $\mu_{ij}$$^d$      & 0.98 & 6e-04 D & \cite{Peterson84} \\
       \hline
       \multicolumn{4}{l}{$^a$Line pairs emitted from a common upper level.} \\
       \multicolumn{4}{l}{$^b\Lambda$ doublets. $^c$Sec. \ref{data}, point 2. $^d$See text.} 
    \end{tabular}
    \vspace{10pt}
    \begin{tabular}{c|c|c|c|c|c|c}
    \hline
    \multicolumn{7}{c}{$\mu_{ij}$ (D)} \\
    \hline
       $\Omega$  & $J$ & $v$ & obs$^e$ & calc$^f$ & $\sigma^g$ & $\Delta^h$ \\
       \hline
       0.5 & 0.5  &  0  & 0.5516 &  0.5516  & 7e-04  &   0.0 \\
       1.5 & 1.5  &  0  & 0.9730 &  0.9730  & 6e-05  &   0.9 \\
       1.5 & 1.5  &  1  & 0.9785 &  0.9786  & 9e-05  &   1.2 \\
       1.5 & 1.5  &  2  & 0.9811 &  0.9823  & 1e-03  &   1.2 \\
       \hline
       \multicolumn{7}{l}{$^e$Recalculated from permanent dipole \cite[Table 3]{Peterson84}. }\\
       \multicolumn{7}{l}{$^f$TDM for the 0-0Q($J$) lines.}\\
       \multicolumn{7}{l}{$^g$Experimental uncertainty.}\\
       \multicolumn{7}{l}{$^h\Delta=\left|\textrm{obs-calc}\right|/\sigma$.}
    \end{tabular}
    \label{tab_fitres}
\end{table}

The fitting results for the relative intensities and permanent-dipole data used in the fitting (see Sec. \ref{data}) are shown in Table \ref{tab_fitres}.  
Permanent dipoles, $\mu$, were calculated by Peterson \etal\ from their measured values of the TDM, $\mu_{ij}$, using theoretical data from other sources. Using the data of \cite[Table 3]{Peterson84}, we restored the measured values of $\mu_{ij}$ and compared them with our calculations. The fit quality is characterized by the root-mean-square (standard) deviation and $\chi_\textrm{red}$.

The results are satisfactory for all data except for the $\Lambda$ doublets, where $\chi_\textrm{red}$ is greater than unity. We remind that the theory of \Brooke\ predicted nearly equal intensities for the components of the $\Lambda$ doublets, and this might seem natural because of very small splitting due to interaction with the remote $A$ state. In fact, however, this is in contradiction with experiment, and the present theory allows differences up to 50\% due to a more adequate account of the $X$-$A$ coupling; namely, the TDMs for the doublet components have different forms. For instance, in Eqs. (\ref{TDMQfeef}) and (\ref{TDMQeffe}) for the Q branch, the second terms in the square brackets have opposite signs. Nevertheless, the problem has not been resolved in full, as testified by rather large $\chi_\textrm{red}$.

All fitting results are given in Supplementary material.

\section{Population differences}\label{PopDif}

\begin{figure}[ht]
    \centering
    \includegraphics[scale=0.4]{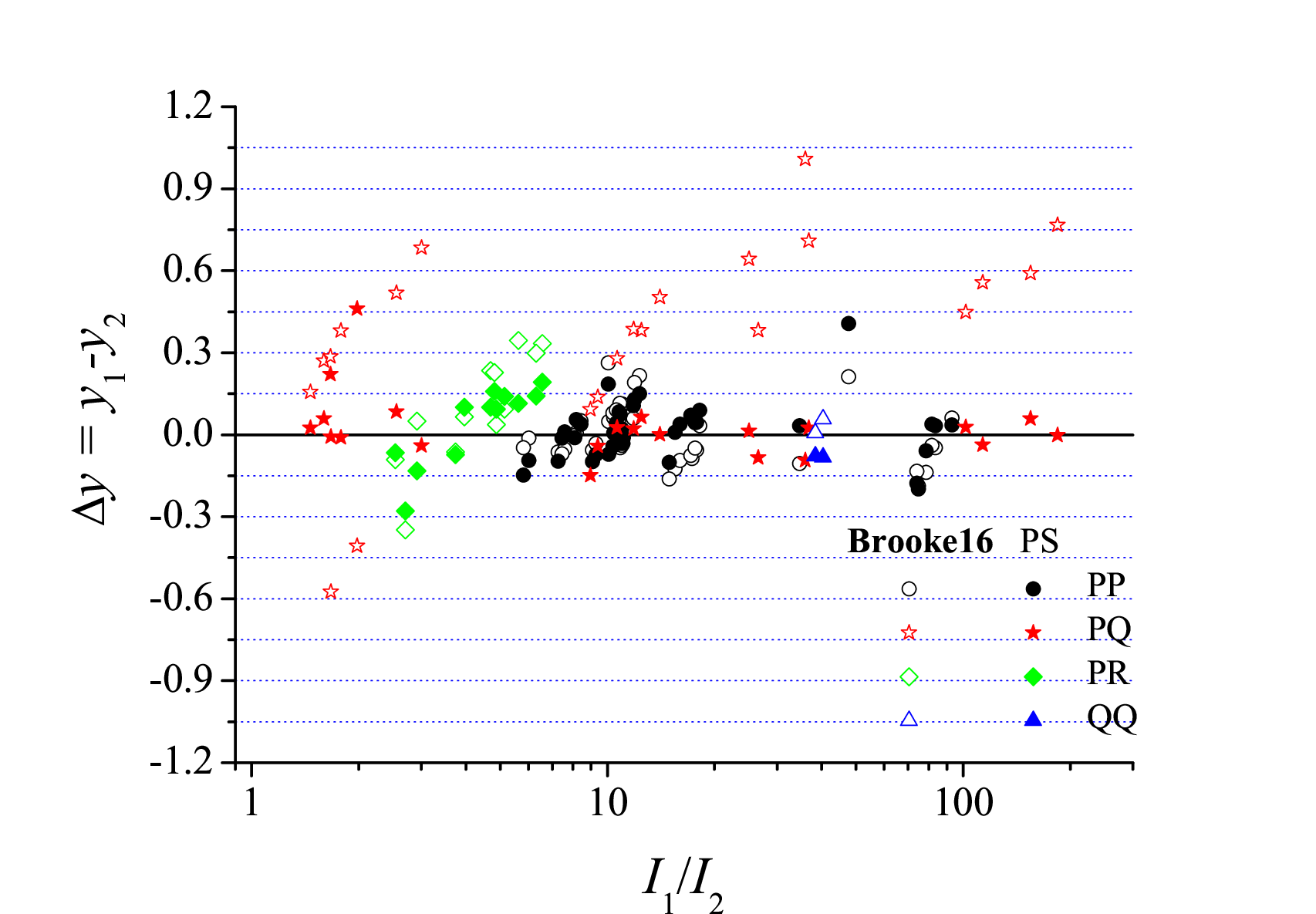}
    \caption{The difference of the logarithmic populations calculated by Eq. (\ref{y}) from the \textbf{Noll20} line intensities for the line pairs emitted from a common upper level. For the calculation, theoretical values of the \EinA\ of the present study (PS) and \textbf{Brooke16} were used. 
    The members of pairs were assigned numbers 1 and 2 such that $I_1/I_2>1$. Empty symbols, \textbf{Brooke16}, $\left<\Delta y\right>=0.12$, std = 0.23; filled symbols, PS, $\left<\Delta y\right>=0.002\approx0$, std = 0.11. The majority of the PS points are within $0\pm0.15$.}
    \label{fig_de}
\end{figure}

The logarithmic populations of the emitting energy levels calculated by Eq. (\ref{y}) should be independent of the line observed, but in fact there is some scatter characterized by the difference, $\Delta y=y_1-y_2$, between the populations calculated from the intensities of lines 1 and 2 emitted from a common upper level. In the \textbf{Noll20} data, there are 114 groups of lines with quality flag 33, each containing two to nine transitions from a common upper level. For analysis, we selected only transitions with resolved $\Lambda$ doublets, a total of 80 pairs of lines. The $\Delta y$ values calculated with the \EinA\ from \textbf{Broke16} and from the present study are shown in Fig. \ref{fig_de}. The stronger lines in the pairs were assigned no. 1, so that the intensity ratios are always greater than unity, $I_1/I_2>1$.

Inspection of Fig. \ref{fig_de} shows that the open symbols (the \Brooke\ data) are appreciably shifted to the positive values whereas the filled ones (PS) are symmetric with respect to the abscissa axis. In figures, if we neglect the 3-5 largest positive and negative outliers, then $\Delta y$ is between $-$0.15 and +0.6 for the former whereas the latter is between $-$0.15 and +0.15. The $\pm$ asymmetry in the \textbf{Brooke16} data might indicate the presence of a systematic error.

The data of Fig. \ref{fig_de} can be characterized by the average values, $\left<\Delta y\right>$, and standard deviations, std. 
The above-noted asymmetry is expressed in these quantities as follows: $\left<\Delta y\right>=0.12$, std = 0.23 for the \Brooke\ data, $0.002(\approx0)$ and 0.11 for the PS. Since the PS $\left<\Delta y\right>$ is negligible as compared to the std, the $\Delta y$ scatter is fully random, as should be in the absence of systematic errors. 
In contrast, the \Brooke\ data seem to contain systematic errors since $\left<\Delta y\right>$ is appreciably non-zero.

Further, the error of the \Noll\ intensities estimated in Ref. \cite{Noll20} is about 10\%, hence the $I_1/I_2$ ratio bears the error of about 15\%. This error should result in $\Delta y$ scatter of about $\pm0.15$.
Considering Fig. \ref{fig_de}, we see that the PS ratios of the \EinA\ indeed provide for the variations in $\Delta y$ within the experimental errors whereas those of \Brooke\ do not.

More information can be obtained from the \EinA\ ratios if branches of lines are specified. There are four types of line pairs marked by different symbols in Fig. \ref{fig_de}. The P lines are  relatively strong, and the PP pairs (circles) are well described by both the theories: except for a few outliers, $\Delta y$ do not go out the experimental error. Among the PR pairs (rhombs), more \Brooke\ points decline from experiment appreciably. The most dramatic disagreement is for the PQ pairs (stars) due to weak Q lines corresponding to high-$\Delta v$ transitions, where the present theory works much better: the filled stars are all within the experimental error. 

In order to quantify the above findings, let 1 be a common upper level, 2 and 3 different lower levels, and consider the ratios, $x=A_{12}/A_{13}$, for transitions $1\rightarrow2$ and $1\rightarrow3$ belonging to given branches. The observed value for the $i$th ratio, $x_i^\textrm{obs}$, can be obtained from the above-mentioned \Noll\ pairs of lines, and then compared with the calculated value, $x_i^{\textrm{calc}}$. To characterize the comparisons, we introduce\footnote{This quantity is also used to characterize comparisons for other parameters.}
\begin{equation}
    \chi_\textrm{red} = \sqrt{\frac{1}{N} {\sum_{i=1}^N}\left(\frac{x_i^{\textrm{obs}}-x_i^{\textrm{calc}}}{\sigma_i}\right)^2}, \label{chi}
\end{equation}
 where summation is over all pairs belonging to a given type, say, the PQ type (\emph{i.e.} $A_{12}$ is for a P line and $A_{13}$ for the corresponding Q line), and $\sigma_i=0.15$ as cited above. The values of $\chi_\textrm{red}>1$ mean that the model disagrees with the data beyond the experimental errors whereas $\chi_\textrm{red}<1$ means that agreement is satisfactory but the errors are overestimated.
\begin{table}[htbp]
    \centering
    \caption{Values of $\chi_\textrm{red}$ for line pairs shown in Fig. \ref{fig_de}}
    \vspace{10pt}
    \begin{tabular}{c|c|c|c}
    \hline
    branches & no. of pairs & \Brooke\ & PS\\
    \hline
     PP    & 44 & 0.7 & 0.7\\
     PQ    & 22 & 2.7 & 0.8\\
     PR    & 12 & 1.4 & 0.9\\
     QQ    &  2 & 0.3 & 0.5\\
     \hline
     total & 80 & 1.62 & 0.77\\
     \hline
    \end{tabular}
    \label{tab_chi_red}
\end{table}

Table \ref{tab_chi_red} shows the results. The most numerous PP pairs of the P-branch lines belonging to strong low-$\Delta v$ transitions are described by both theories equally well. In contrast, the R and Q lines are relatively weak and the \Brooke\ theory gives the errors in the PR and PQ pairs well beyond the experimental error whereas the present theory agrees with the observations. The number of the QQ pairs is insufficient for definite judgement. 

The present theory is expected to work better at higher $J$. Table \ref{tab_highJ} shows four PQ pairs with $J^\prime=5.5$ and 6.5 found in the \Noll\ data.
\begin{table}[htbp]
    \centering
    \caption{The ratios of the \EinA\ for some PQ pairs}
    \vspace{10pt}
    \begin{tabular}{c|c|c|c|c|c|c|c|c|c|}
       branch & $v^\prime$ & $J^\prime$ & $p^\prime$ & $v^{\prime\prime}$ & $J^{\prime\prime}$ & $p^{\prime\prime}$ & \Noll & \Brooke & PS \\
       \hline
       P & 7 & 5.5 & e & 2 & 6.5 & e &  36.0$^a$   &  13.1$^a$   &  39.5$^a$  \\
       Q & 7 & 5.5 & e & 2 & 5.5 & f &             &  $2.7^b$    &  $0.9^b$  \\
       P & 7 & 5.5 & f & 2 & 6.5 & f &  25.0$^a$   &  13.2$^a$   &  24.7$^a$  \\
       Q & 7 & 5.5 & f & 2 & 5.5 & e &             &  $1.9^b$    &  $1.0^b$  \\
       P & 7 & 6.5 & f & 3 & 7.5 & f &  26.6$^a$   &  18.1$^a$   &  28.8$^a$  \\
       Q & 7 & 6.5 & f & 3 & 6.5 & e &             &  $1.5^b$   &  $0.9^b$  \\
       P & 7 & 6.5 & e & 3 & 7.5 & e &  36.8$^a$   &  18.1$^a$   &  36.7$^a$  \\
       Q & 7 & 6.5 & e & 3 & 6.5 & f &             &  $2.0^b$   &  $1.0^b$  \\
       \hline
       \multicolumn{10}{l}{$^a$ The observed, $x_i^\textrm{obs}$, and calculated, $x_i^\textrm{calc}$, P/Q ratios.} \\
       \multicolumn{10}{l}{$^b$ $x_i^\textrm{obs}/x_i^\textrm{calc}$.}
    \end{tabular}
    \label{tab_highJ}
\end{table}
The ratios shown are well reproduced by the present study whereas the \Brooke\ one fails. A total of 22 such pairs are present in \Noll, they all are given in Supplementary material.

It is interesting to note that twenty two PQ pairs are emitted from eleven $\Lambda$ doublets.
The \Brooke\ theory gives nearly identical \EinA\ for the components of the $\Lambda$ doublets,\footnote{As indicated in \Noll, the maximum difference is 0.25\%.} examples are seen in Table \ref{tab_highJ} where four upper states form two doublets, for which \Brooke\ gives equal ratios whereas the observation gives differences about 30\% well reproduced by the present study. In general, the present theory predicts differences up to 50\% for high $\Delta v$ and $J$.

\begin{figure}[ht]
    \centering
    \includegraphics[scale=0.2]{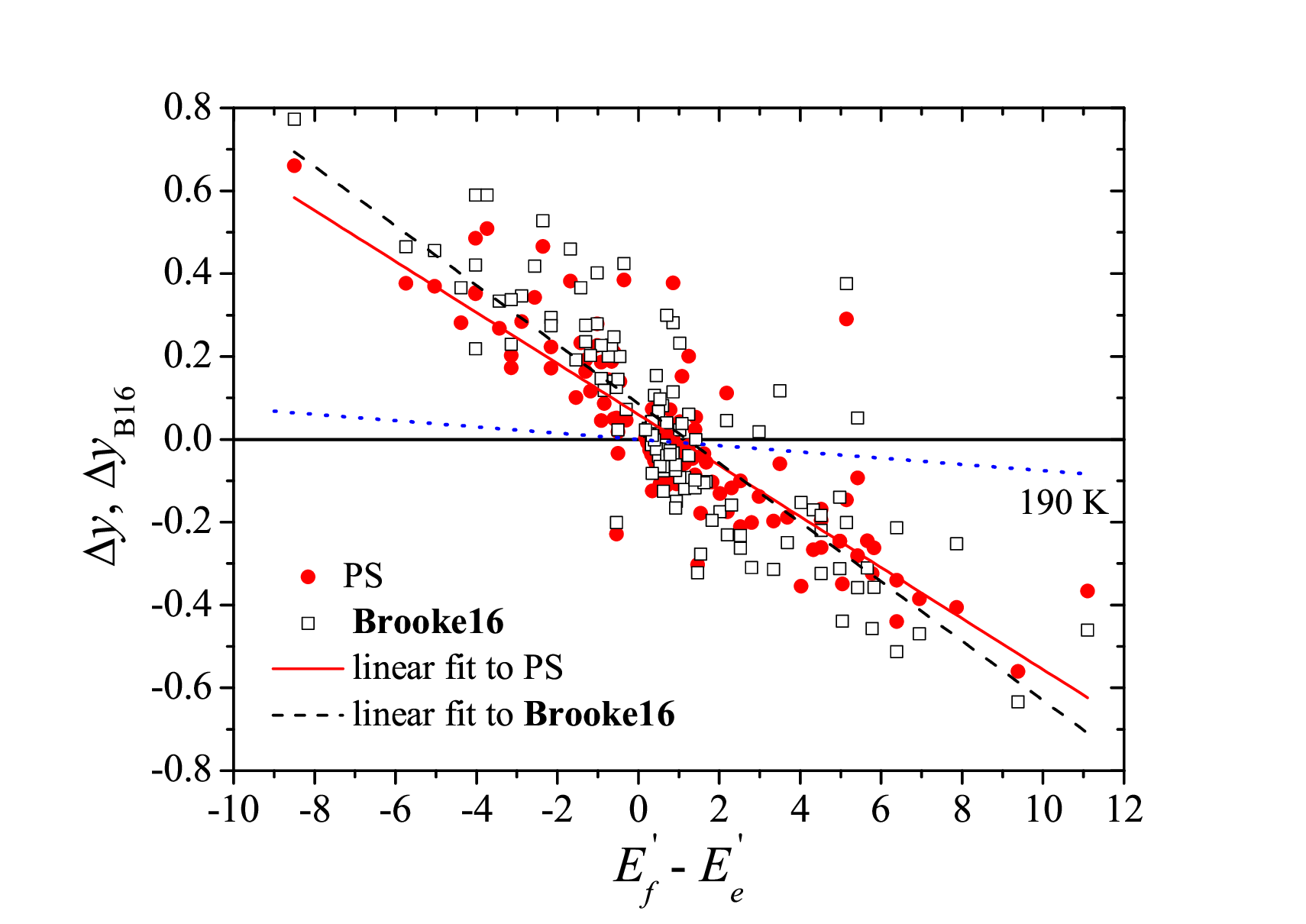}
    \includegraphics[scale=0.2]{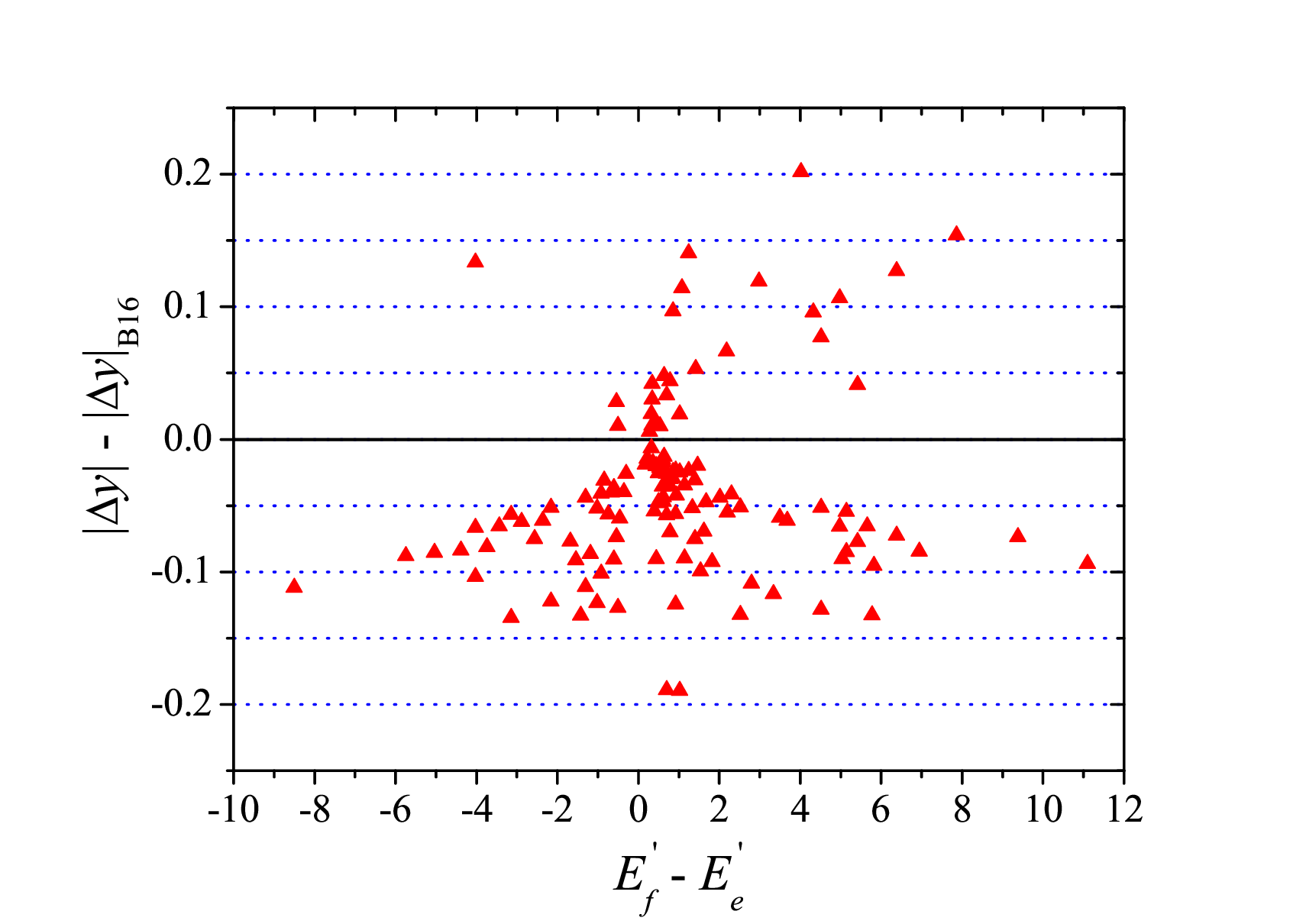}
    \caption{The difference of the logarithmic populations between the components of the $\Lambda$ doublets calculated from the \textbf{Noll20} line intensities with use of the \EinA\ of the present study, $\Delta y$, and with the \textbf{Brooke16} coefficients, $\Delta y_\textrm{B16}$.}
    \label{figdy}
\end{figure}

In continuation of the above subject, we consider the population difference between components of the $\Lambda$ doublets as function of the doublet splitting shown in two panels of Fig. \ref{figdy}. If local thermodynamic equilibrium (LTE) exists in the emitting region, the slope of the linear fit permits evaluation of the local temperature. 

The left panel of Fig. \ref{figdy} is similar to Fig. 9 of \Noll. The dotted line shows the expected distribution for LTE at 190 K. The actual distributions predicted by the two theories are far from this expectation. While the present study provides for some improvement, it remains unsatisfactory and requires further work.

The above-mentioned improvement is quantified on the right panel. There are 27 points with positive ordinate, where the \Brooke\ works better, and 101 points where better is the present study.

As is obvious from the above considerations, the problem of population differences discussed in this section is closely related to the precision of the theoretical calculations, as scrutinized in \textbf{Noll20}, where some inconsistencies in the theoretical data are emphasized.

In particular, 
French \emph{et al.} \cite{French00} measured the Q$_1/$P$_1$ and R$_1/$P$_1$ airglow emission ratios in the 6-2 band and found them to be lower than those predicted by the theory \cite{Mies74,Langhoff84,Turnbull89}. Pendleton and Taylor \cite{Pendleton02} explained this disparity as being due to the $A$-state perturbation. In Table \ref{FrenchTable}, we compare the experimental ratios with the ones of \textbf{Brooke16} and of the present study (column PS). All the observed ratios are better described by the present theory, the difference with the observation mostly being within the experimental error. Moreover, if we remove state $A$ from the model by putting $d^A(r)$ and $d^{XA}(r)$ identically equal to zero, our ratios become nearly coinciding with those of \textbf{Brooke16}, as shown in column PS0. In the next section, we repeat this test with $X$-$A$ uncoupling and find the same result, see Fig. \ref{figNIDLstrong}. Thus, proper account of the $A$ state indeed has positive effect on the intensities.
\begin{table}[htbp]
    \centering
    \caption{Comparison of the observed and calculated Einstein $A$ coefficients for the 6-2 band}
    \vspace{10pt}
    \begin{tabular}{c|c|c|c|c|c|c}
    \hline
     $J^{\prime}$ & Ratio  & \textbf{Brooke16} & PS0$^a$ & PS &  Experiment \cite{French00} & $\Delta^b$ \\
     \hline
      1.5   & $Q_1(1)/P_1(2)^c$ & 1.322 & 1.324 & 1.267  & 1.261$\pm$0.012 & 0.5\\
      2.5   & $Q_1(2)/P_1(3)$ & 0.441 & 0.443 & 0.398  & 0.389$\pm$0.006 & 1.5\\
      3.5   & $Q_1(3)/P_1(4)$ & 0.215 & 0.216 & 0.178  & 0.173$\pm$0.011 & 0.5\\
      4.5   & $Q_1(4)/P_1(5)$ & 0.124 & 0.125 & 0.091 & 0.094$\pm$0.004 & 0.7\\
      2.5   & $R_1(1)/P_1(3)$ & 0.451 & 0.453 & 0.445 & 0.436$\pm$0.017 & 0.5\\
      3.5   & $R_1(2)/P_1(4)$ & 0.532 & 0.535 & 0.522 & 0.510$\pm$0.010 & 1.2\\
      4.5   & $R_1(3)/P_1(5)$ & 0.540 & 0.544 & 0.527 & 0.483$\pm$0.018 & 2.4\\
      \hline
      \multicolumn{6}{l}{$^a$Present study with the $A$ state removed. $^b\Delta=\left|\textrm{obs-calc}\right|$/err.}\\
      \multicolumn{6}{l}{$^c$Line notations as in Ref. \cite{French00} (values of $N^{\prime\prime}=J^{\prime\prime}-\tfrac{1}{2}$ }\\
      \multicolumn{7}{l}{\hspace{5pt}{are indicated in braces).}}
      \end{tabular}
      \label{FrenchTable}
\end{table}

Franzen \emph{et al.} \cite{Franzen19} observed several $\Delta v=2$ and 3 bands in airglow not covered by the \textbf{Noll20} data. The measured line intensities were converted into the Einstein $A$ coefficients of four Q lines using the observed relative Q/P intensities and the theoretical (HITRAN) coefficients for the P lines, which were considered reliable. Unfortunately, the raw intensity data were not published, therefore we cannot make a comparison similar to Table \ref{FrenchTable}.

We conclude that the present theory is essential improvement of the Einstein $A$ coefficients and the derived level populations, though unresolved issues remain.

\section{Intensity distributions} \label{NIDLs}

We have already analyzed the intensity distributions in terms of the Normal Intensity Distribution Law (NIDL) \cite{Medvedev24MP} and found, in particular, that the NIDL is not fulfilled for the satellite lines. Here, we consider the intensity distributions in more detail.

\begin{figure}[htbp]
    \centering
    \includegraphics[scale=0.2]{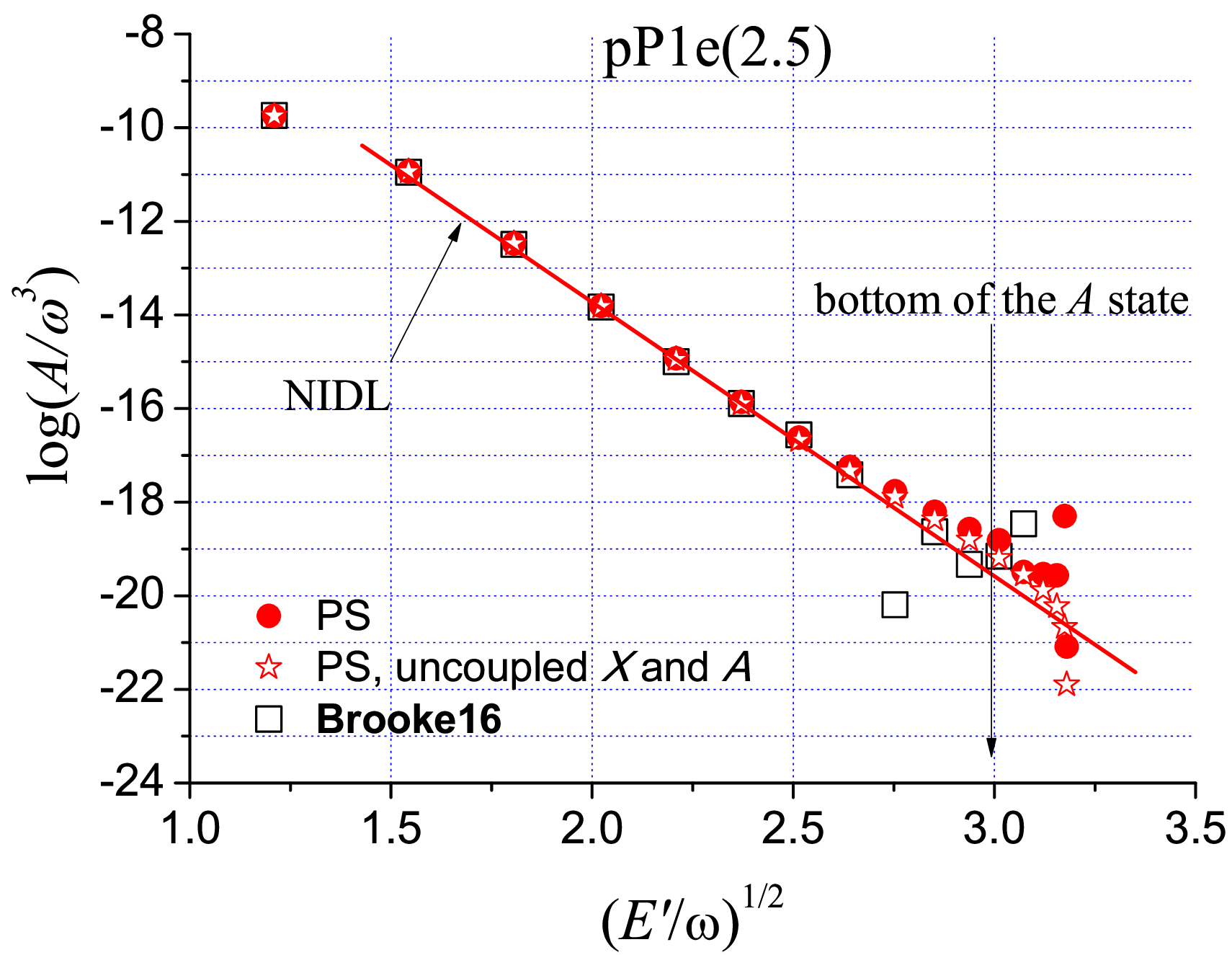}
    \includegraphics[scale=0.2]{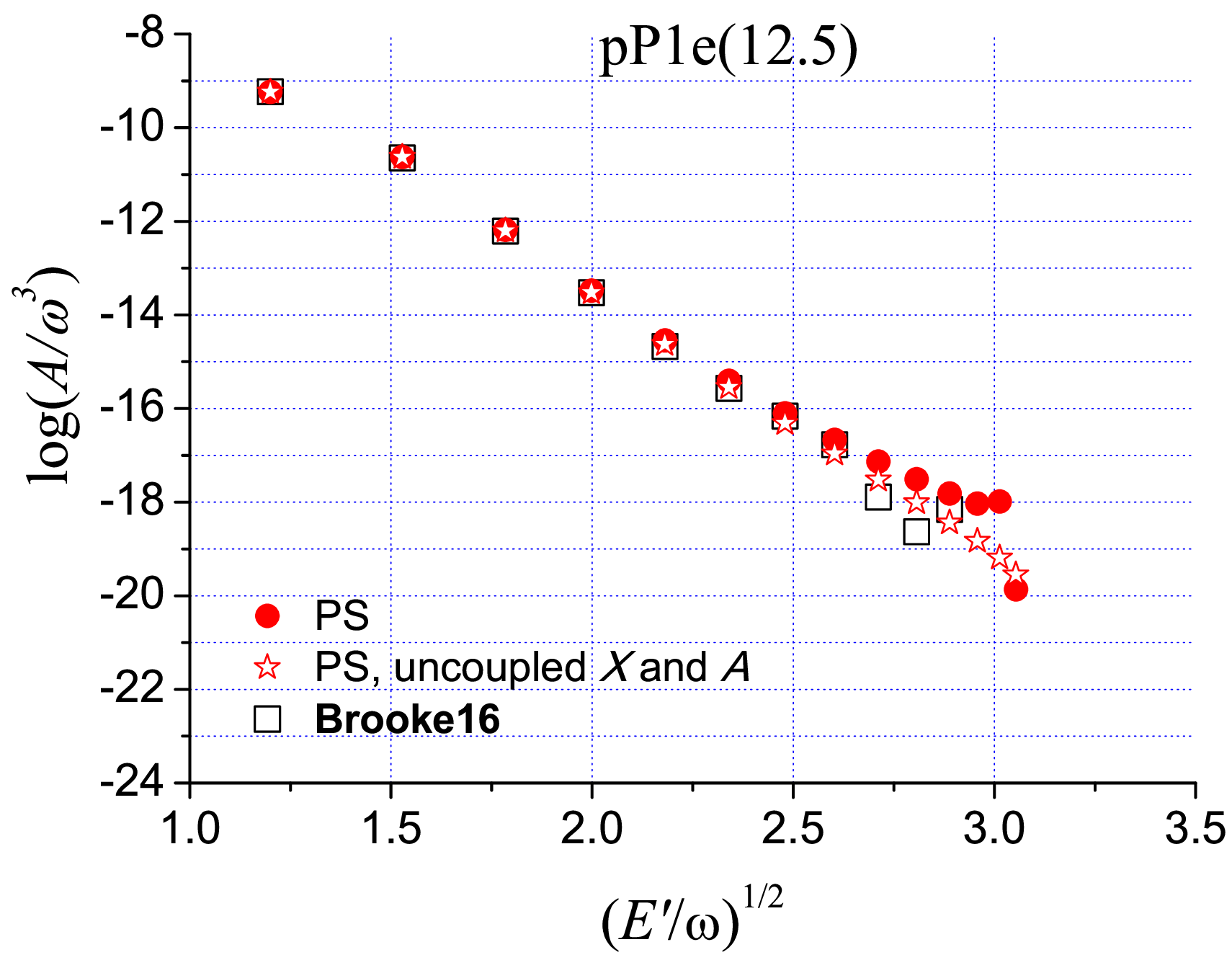}
    \includegraphics[scale=0.2]{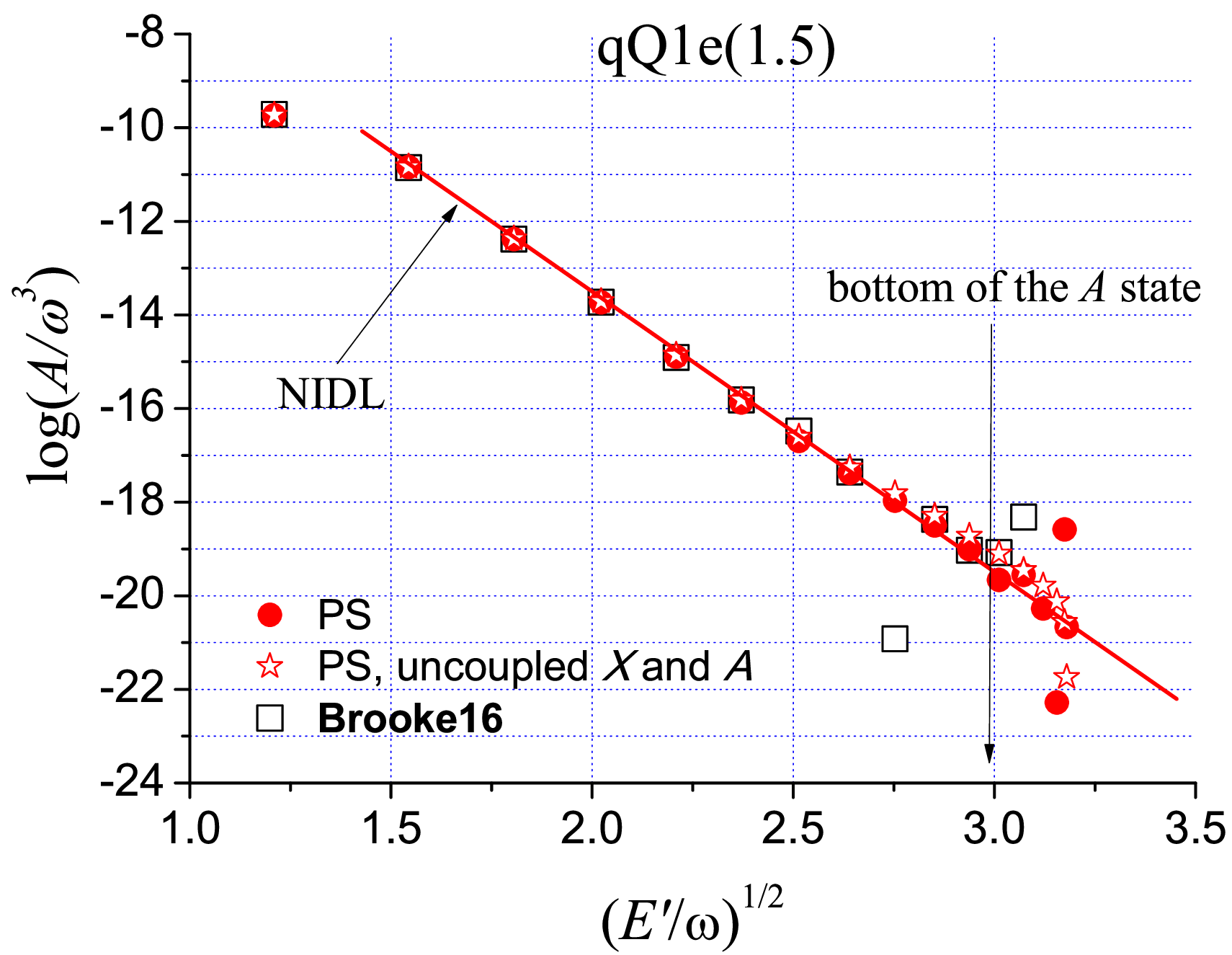}
    \includegraphics[scale=0.2]{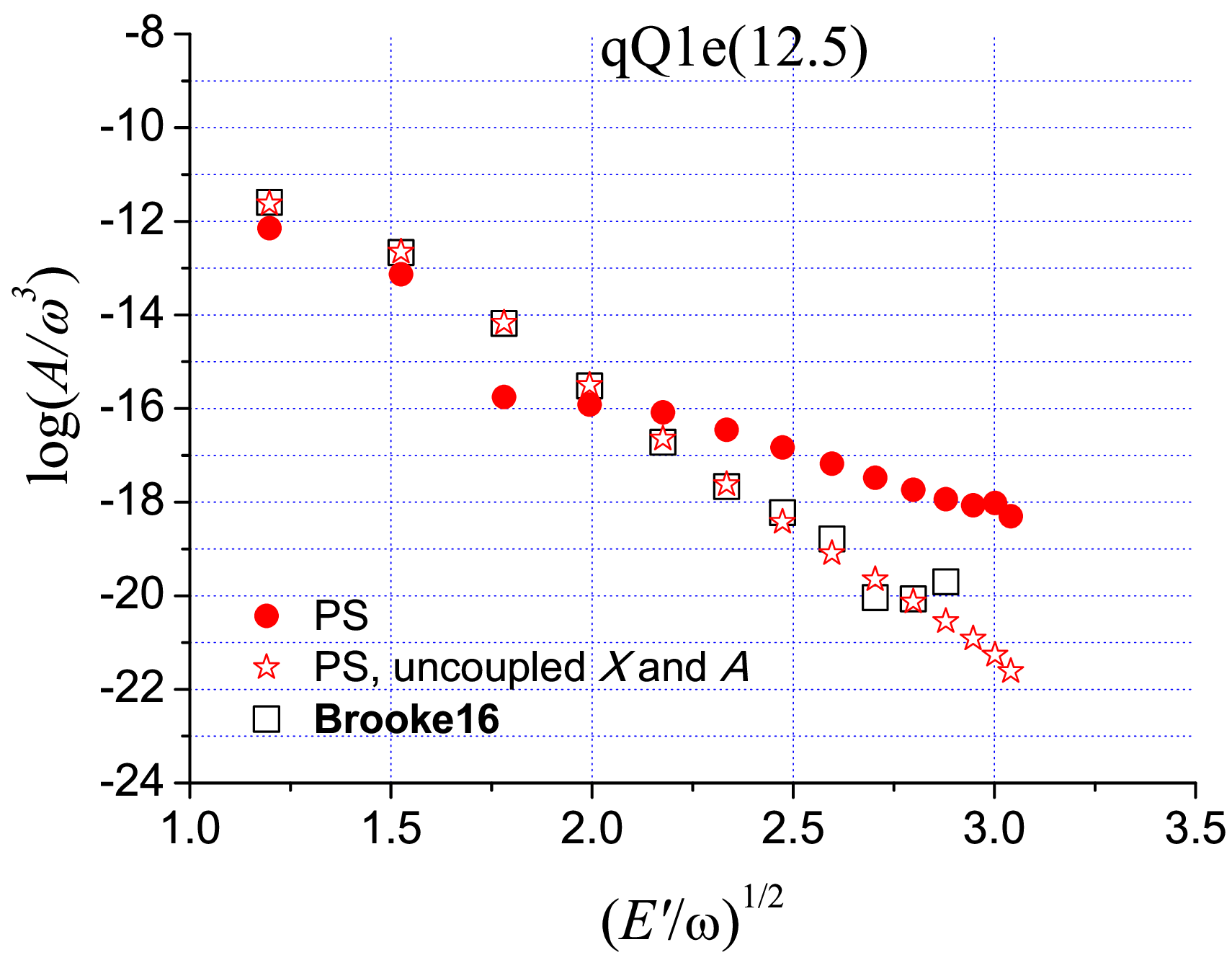}
    \includegraphics[scale=0.2]{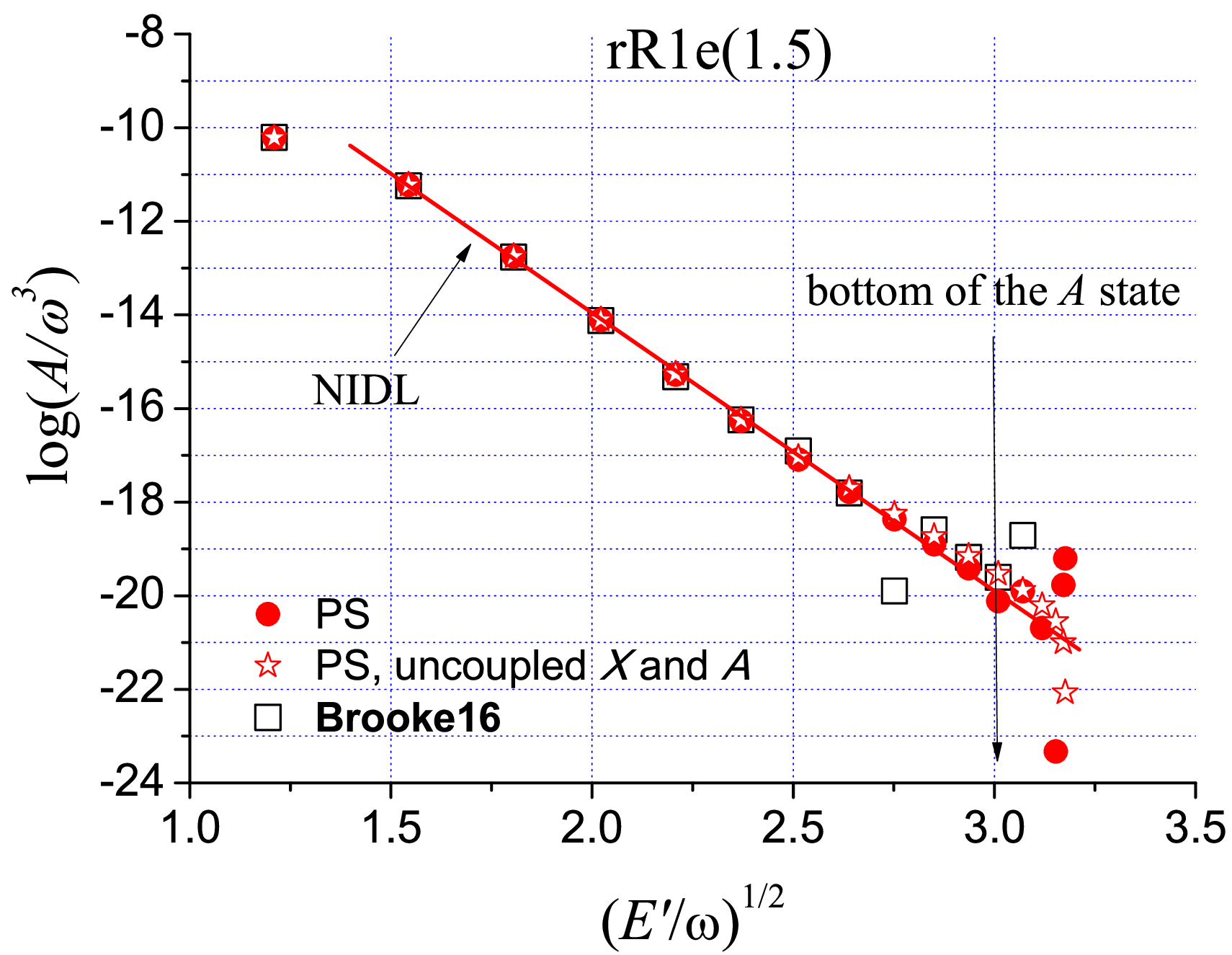}
    \includegraphics[scale=0.2]{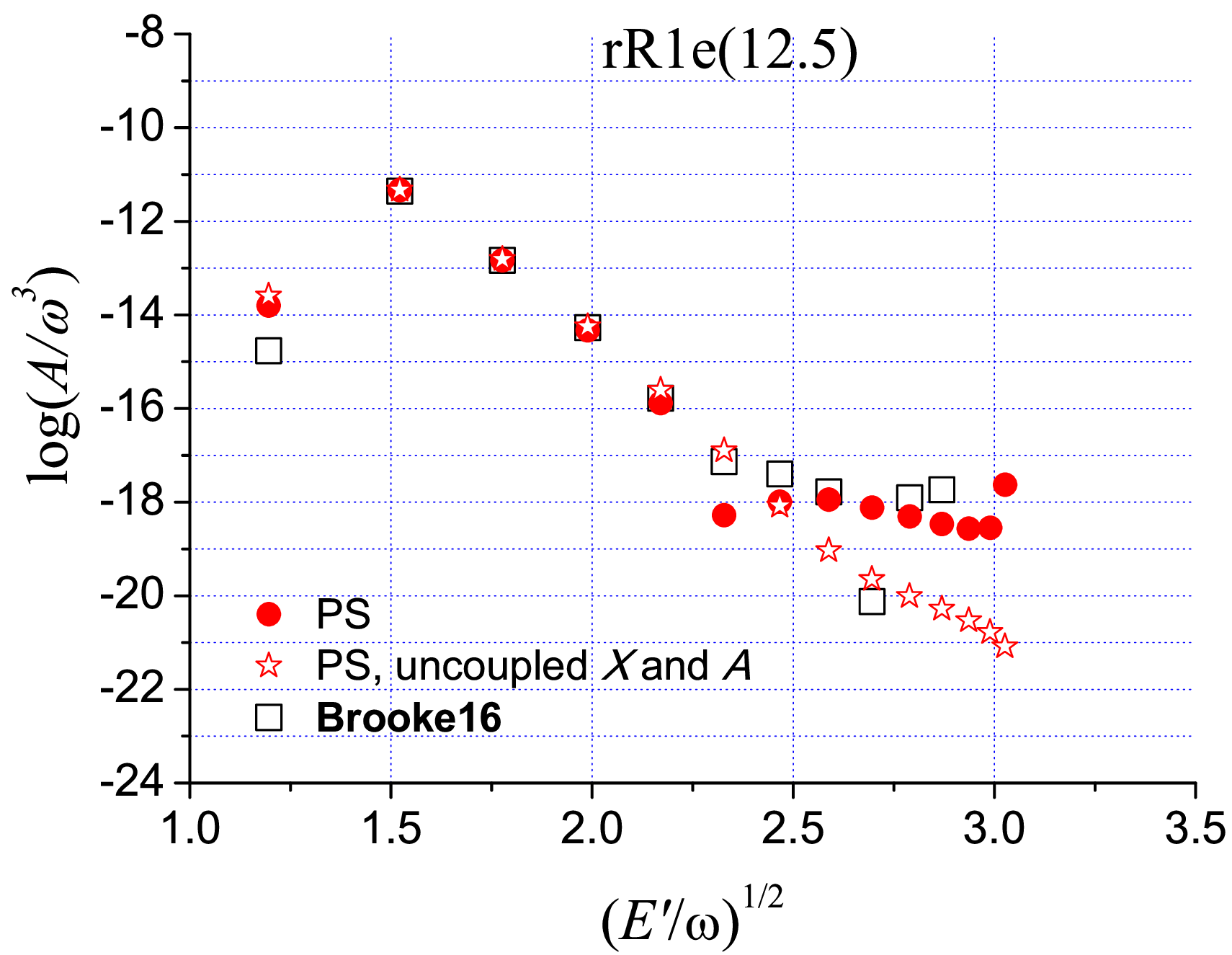}
    \caption{Intensities of the $v$-0 ($v=1$-17) main transitions in the NIDL coordinates. Line notations as in \Brooke\ (values of $J^{\prime\prime}$ are indicated in braces). $A$ in s$^{-1}$, $E^\prime$ is the upper-state energy in cm$^{-1}$, $\omega=3738$ cm$^{-1}$ \cite{Brown79} is the harmonic frequency. The NIDL line is drawn over $v=2$-12 points. The bottom of the $A$ state is between $v=11$ and 12; the NIDL is not obeyed at $v>12$ where jumps are present due to resonances with the $A$-state levels. Anomalies due to destructive interference \cite{Medvedev12,Ushakov24} are seen, \emph{e.g.} lines 1-0R1e(12.5), 9-0qQ1e(1.5), \emph{etc.} in the \Brooke\ data.}
    \label{figNIDLstrong}
\end{figure}

The distributions for the strong lines due to the main transitions are shown in Fig. \ref{figNIDLstrong}. Each panel contains three sets of data: present (PS, circles) and previous (\Brooke, squares) studies as well as PS with a truncated model (stars), in which the interaction between the $X$ and $A$ states is turned off, so that the system of two equations remains to solve. 
Comparing these data permits evaluation the effect of the $A$ state on the intensities of the $X$-$X$ transitions.

The left and right panels present results for the low- and high-$J$ transitions, respectively. We will discuss them in turn.

Consider first the PS data on the \underline{left panels} (circles). The $X$-$A$ coupling is small; it manifests only at sufficiently large $v$, $v>7$, when the upper vibrational level in the $X$ state approaches the bottom of the $A$ state (compare circles with stars). The effect of the $A$ state becomes especially strong at abscissa greater than 3.0 where resonances arise with the $A$ vibrational levels. 

The \Brooke\ data (squares) mostly follow our model at $v\le7$. The jumps at $v=8$ or 9 in these data
 are due to discontinuity in the third derivative of the potential noted in Ref. \cite{Medvedev24MP}. With the analytic PEF of the present study, these jumps disappear, yet another jumps at $v>12$ arise due to the above-noted resonances. These latter jumps also disappear when the $X$-$A$ coupling is switched off. 

The NIDL theory was developed for the $X$-$X$ transitions between the purely vibrational states within a single potential. In the present model, there are two effective potentials in state $X$ that are only slightly modified by rotation and are nearly parallel to each other (\emph{i.e.} they have nearly identical vibrational wave functions and repulsive branches); therefore, the NIDL theory is expected to apply until the $X$-$A$ coupling remains insignificant. Indeed, it is seen that the NIDL theory works well at $v=2$-7 and the slope of the NIDL is essentially the same for all branches, which results in similar intensities. The $X$-$A$ coupling begins to affect the intensities at $v\ge8$, where the intensities increase with respect to the NIDL.

In the \underline{right panels}, we do not draw the NIDL lines because no NIDL theory has been developed for the case where rotation comes into play. Nevertheless,
the NIDL seems to be fulfilled at $v=2$-8 for all branches when the $A$ state is turned off (stars). 

At high $J$, the effect of the $A$ state becomes dominant and is drastically different between the branches. The centrifugal potential strongly affects the energy levels and the steepness of the repulsive branch, which determines the rate of intensity fall-off with increasing $v$. This results in different rates with which various terms in TDMs decrease with increasing $v$. The terms with superscript $X$, which provide the largest contributions
in Eqs. (\ref{TDMPbranch})-(\ref{TDMReeff}) at small $v$, decay with $v$ most rapidly in the R branch and least rapidly in the P branch, as illustrated by stars in the right panels. Thus, at $v=2$ (abscissa 1.5), the intensities of the P and R lines differ only a few times, while at $v=7$ (abscissa 2.5) the R line is weaker by about two orders of magnitude.

The $X$ terms are largest only at small $v$. When $v$ increases, the relative contributions of the $XA$ terms also increase; they become comparable with $X$ at $v=6$ and then continue to increase as a consequence of approaching the minimum of the $A$-state potential mentioned above. As a result, the coupling with the $A$ state becomes the major factor; therefore, at abscissa 3.0, the line intensities become approximately the same in all three branches. The effect of the coupling being larger in R than in P is due to the fact mentioned above that the $X$ terms in R are much less than in P.

The largest effect of the $A$ state is observed in the Q branch. First of all, it is seen that circles are shifted with respect to stars and squares at all $v$. In other words, our intensities are essentially different from those of Brooke \etal\

Second, there is an effect specific for the Q branch. Namely, 
the $X$ and $A$ terms in the Q-branch TDMs are divided by $J$ (see Eqs. (\ref{TDMQfeef}) and (\ref{TDMQeffe})), hence, their contributions to the low-$v$ intensities are inversely proportional to $J^2$ \cite{Mies74} being hundreds of times less than in P and R. In the absence of the $A$ state (stars and squares), this persists at high $v$ as well.
The coupling with the $A$ state increases the intensities cardinally since the $XA$ terms in the TDM survive at large $J$ whereas the $X$ and $A$ terms go off. Therefore, at abscissa 3.0, the Q lines are as intense as P and R.

Note that the \Brooke\ points follow the truncated model very closely, therefore the TDMs for the Q branch at $v\ge5$ are very underestimated.

\begin{figure}[htbp]
    \centering
    \includegraphics[scale=0.2]{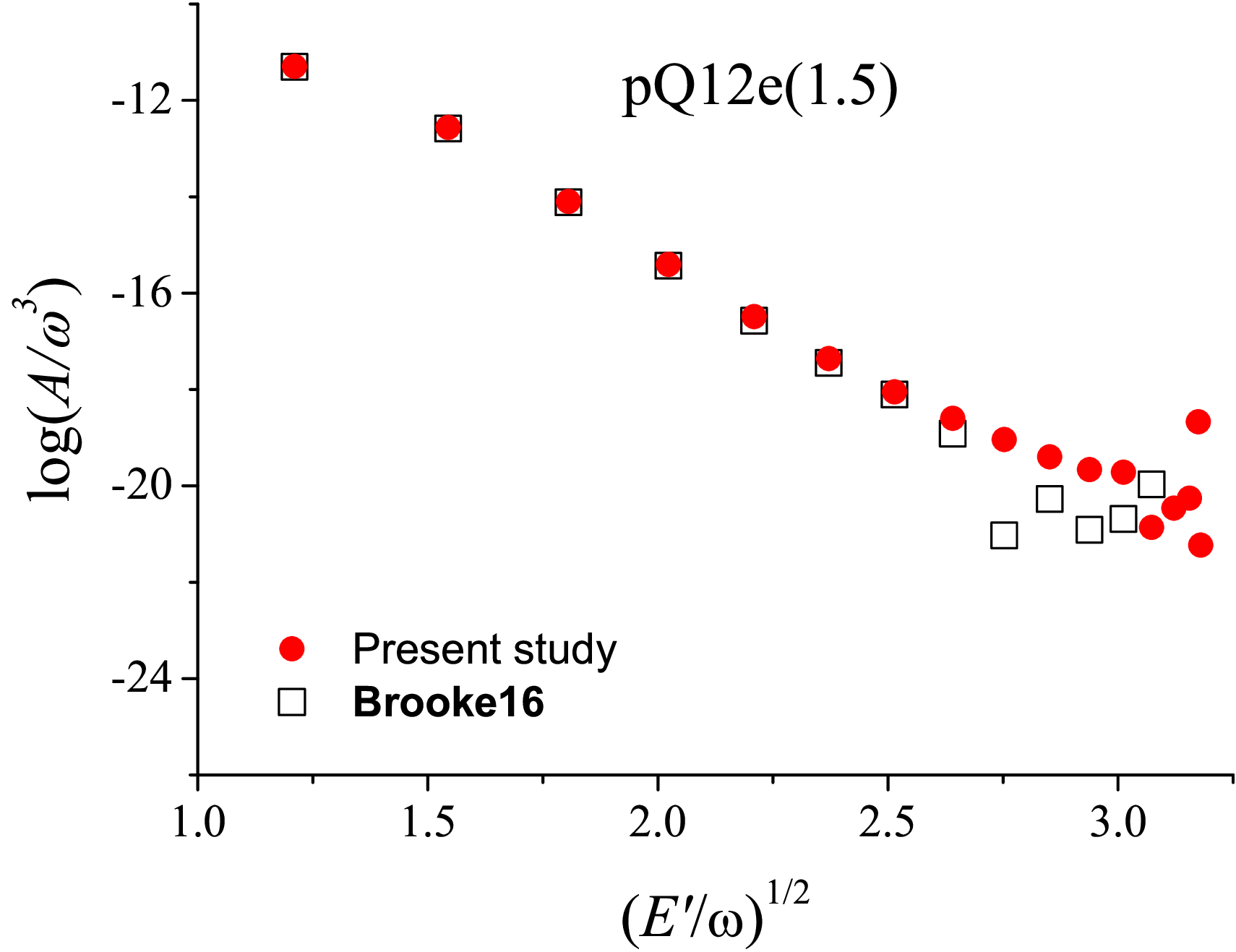}
    \includegraphics[scale=0.2]{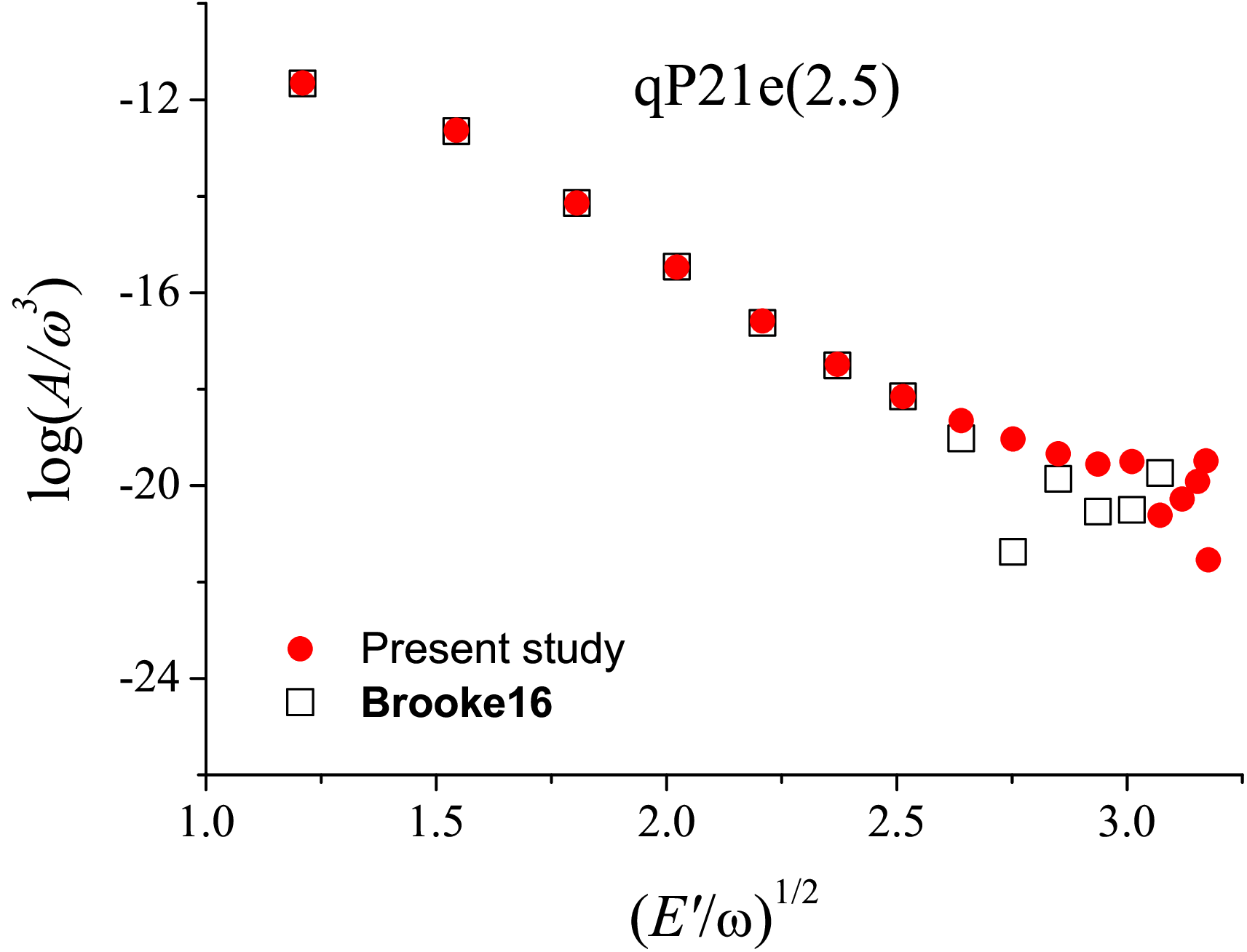}
    \includegraphics[scale=0.2]{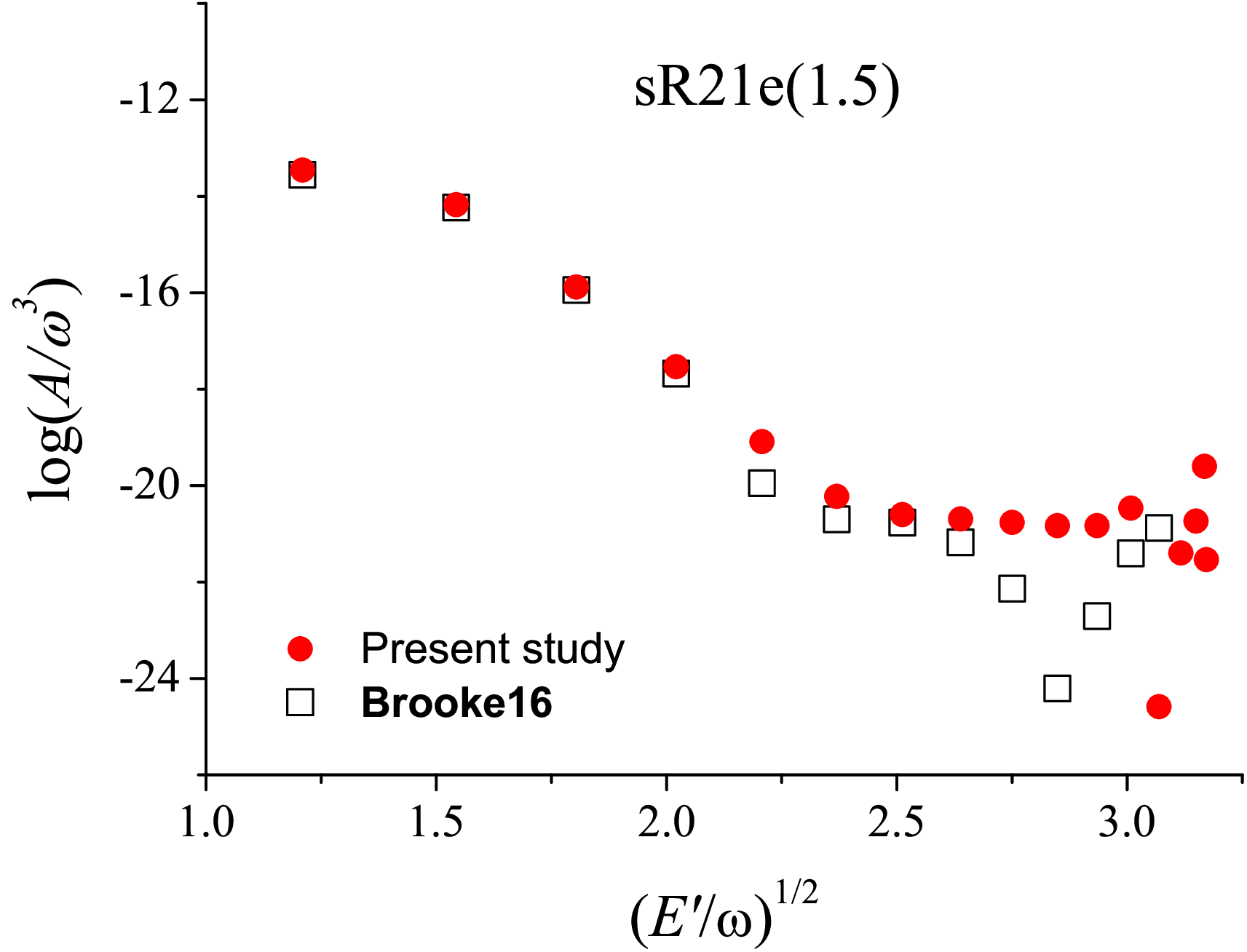}
    \caption{Intensities of the $v$-0 ($v=1$-17) satellite transitions in the NIDL coordinates.}
    \label{figNIDLsR21e25}
\end{figure}

The distributions for the weak lines due to the satellite transitions are shown in Fig. \ref{figNIDLsR21e25}. The figure demonstrates the breakdown of the NIDL \cite{Medvedev24MP}, which is especially obvious for the sR lines. For these lines, essential differences with \Brooke\ are seen at all transitions.

The pQ and qP lines seemingly follow straight lines for the first five overtones, but this should be considered as occasional phenomenon because no theory has been developed for transitions involving change in the potential. Appreciable differences with \Brooke\ take place at $v>5$.

\section{Conclusions}\label{Concl}

The main objective of the present study was correct account of the coupling between the $X^2\Pi$ ground and $A^2\Sigma^+$ excited electronic states. The model used permitted to decrease the scatter (down to the experimental error) of the logarithmic populations of the ro-vibrational levels derived from the observed radiation fluxes, to achieve better agreement with the relative intensities, both measured in laboratory and observed in astronomy, and to obtain large differences in the observed intensities of the $\Lambda$-doublet components, which were not explained by the theory of Brooke \emph{et al.} \cite{Brooke16}. A new line list was calculated (see Supplementary material), which includes the transition frequencies of Brooke \emph{et al.} and the Einstein $A$ coefficients calculated in the present study.

Yet, the present model cannot be considered fully successfull because of the following shortcomings.

1. The differences in the observed and calculated logarithmic populations of the components of the $\Lambda$ doublets shown in Fig. \ref{figdy}, while decreased with respect to the \Brooke\ theory, remains too large to permit correct determination of the local temperature.

2. The energy levels and transition frequencies are reproduced worse than in the Brooke \emph{et al.} theory.

3. Some fitted functions are not smooth, see Fig. \ref{figVcorr} and Fig. S1 in Supplementary material, or strongly decline from \emph{ab initio}.


The solution of the above problems seems to be in proper account of the higher purely repulsive states $1^4\Sigma^-,1^2\Sigma^-$, and $1^4\Pi$ \cite{Mitev24} converging to the $X^2\Pi$ ground-state dissociation limit.

The final remark concerns the experimental data on the intensities of the $X$-$A$ transitions \cite{Luque98}, which could be fitted by the present model. We did not do this, instead we calculated $\chi_\textrm{red}=2.7$ (excluding one outlier). The model of Ref. \cite{Yousefi18} gives $\chi_\textrm{red}=1.9$. Thus, both models do not describe these intensities satisfactorily. The necessity of the model refinements will be dictated by the future experimental demands.

\vspace{20pt}









\section*{Acknowledgements}

We are grateful to P.~F.~Bernath for the data of Ref. \cite{Yousefi18} and S.~Noll for the data of Ref. \cite{Noll20} and helpful discussions. This work was performed under state task, Ministry of Science and Education, Russian Federation, state registration number 124013000760-0.

\section*{Appendix: Why the satellite lines are weak?}\label{app}

In general, the satellite transitions ($\Delta F\ne0$) are weak as compared to the main ones ($\Delta F=0$) because the dipole moment is independent of spin and the satellite lines arise only due to rotational mixing of the multiplet components. However, the rotational perturbation increases with $J$, which requires special consideration.

Mies \cite{Mies74} and Nelson \emph{et al.} \cite{Nelson89a}
explained the weakness of the $\Delta F\ne0$ transitions by small
overlap of the  wave functions in the $F=1$ and 2 electronic $X$
sub-states, and the effect of cancellation of large contributions to
the TDM. It should be noted that the overlap is small only
at low $J$ since the rotational perturbation
becomes on the order of $B_\textrm{e} J\approx100$ cm$^{-1}$ at
$J=5.5$, which is comparable with the multiplet splitting of about 140
cm$^{-1}$. In contrast, the cancellation is more universal and
becomes even more complete with $J$ increasing. Here, we propose a
simple, explicit explanation of the cancellation effect.

Neglecting, for simplicity, the coupling with the $A$ state, we get the
Hamiltonian matrix in Eq. (\ref{Hmatrix2}) as two $2\times2$ 
identical blocks with PEFs $ U_{1}$ and $U_{2}$ (we call them
\textquotedblleft diabatic" states\footnote{Terms ``diabatic, adiabatic, nonadibatic" come from the collision theory.} coupled by perturbation $V$). Let
us diagonalize each block by rotation
over angle $\Theta $ (as in Refs. \cite{Mies74,Nelson89a}) transforming the diabatic functions, $\psi _{1p}$ and $%
\psi _{2p}$, into the \textquotedblleft adiabatic" ones, $\Phi _{+}$ and $\Phi
_{-}$,%
\begin{eqnarray}
\psi _{1p} &=&\Phi _{+}\cos {\Theta }+\Phi _{-}\sin {\Theta ,}  \nonumber \\
\psi _{2p} &=&-\Phi _{+}\sin {\Theta }+\Phi _{-}\cos {\Theta },  \nonumber
\end{eqnarray}%
where $\Phi _{+}$ and $\Phi _{-}$ are solutions of the coupled Schr\"{o}%
dinger equations with the adiabatic potentials, $U_{\pm }=\tfrac{1}{2}\left(
U_{1}+U_{2}\right) \mp \tfrac{1}{2}\sqrt{\left( U_{1}-U_{2}\right)
^{2}+4V^{2}}$. 
The coupling between the $\pm $ states (the \textquotedblleft
nonadiabatic" one) is proportional to derivatives (due to $T_\textrm{vib}$) of the rotation angle %
$\Theta=\frac{1}{2} \arctan{\left(2V/(U_2-U_1)\right)}$. Since $U_2-U_1$ is nearly constant approximately equal to the spin-orbit splitting, $A_X$, and $V$ is increasing with $J$ (see Eqs. (\ref{U1}), (\ref{U2}), and (\ref{V})), this angle, at large $J$, has an almost constant value of $\Theta \approx \pi /4,$ therefore the nonadiabatic coupling can be neglected.
Thus, two types of vibrational states are formed: the states of the
first type are described by the wave functions with component $\Phi
_{-}$ almost equal to zero, whereas, for the second type of states, the
component $\Phi _{+}$ is negligible. The satellite lines correspond
to transitions between states of different types. Let, for
definiteness, the upper and lower vibrational states be
of the first and second types, respectively. Then, the primed and
double-primed diabatic functions can be presented as%
\begin{eqnarray}
\psi _{1p}^{\prime } &=&\Phi _{+}^{\prime }\cos {\Theta ,\quad }\psi
_{2p}^{\prime }=-\Phi _{+}^{\prime }\sin {\Theta ,}  \nonumber \\
\psi _{1p}^{\prime \prime } &=&\Phi _{-}^{\prime \prime }\sin {\Theta
,\quad }\psi _{2p}^{\prime \prime }=\Phi _{-}^{\prime \prime }\cos {\Theta .%
}  \nonumber
\end{eqnarray}%
Calculating $d_{ip^{\prime },kp^{\prime \prime }}^{X}$ by Eq. (\ref{dXdAdXA}), we find for the satellite lines
\[
d_{1p,1p}^{X}=\int_{0}^{\infty }\psi _{1p}^{\prime }d^{X}\psi _{1p}^{\prime
\prime }\ dr=\int_{0}^{\infty }\Phi _{+}^{\prime }\cos {\Theta \ }d^{X}\
\Phi _{-}^{\prime \prime }\sin {\Theta \ }dr
\]%
and
\[
d_{2p,2p}^{X}=\int_{0}^{\infty }\psi _{2p}^{\prime }d^{X}\psi _{2p}^{\prime
\prime }\ dr=-\int_{0}^{\infty }\Phi _{+}^{\prime }\sin {\Theta \ }d^{X}\
\Phi _{-}^{\prime \prime }\cos {\Theta \ }dr. \notag
\]%
We see, that under the ``adiabatic" conditions, \emph{i.e.} at large $J$ (conventionally called Hund's case (b)), the sum of $d_{1p,1p}^{X}$ and $%
d_{2p,2p}^{X}$ vanishes. We will write this in the form of ratio,
\begin{equation}
\frac{d_{2p,2p}^{X}}{d_{1p,1p}^{X}}=-1. \tag{A1}
\end{equation}
As a result, the contributions of $d_{1p,1p}^{X}$
and $d_{2p,2p}^{X}$ into expressions (\ref{TDMPbranch})-(\ref{TDMReeff}) for TDMs of the $P$ and $R$ transitions cancel each other at large $J$. The numerical calculations in
Refs. \cite{Mies74,Nelson89a,Brooke16} showed that the cancellation becomes
more complete with $J$ increasing. 

For strong transitions between states of the first type, we get
\[
d_{1p,1p}^{X}=\int_{0}^{\infty }\cos^2 {\Theta} \ \Phi _{+}^{\prime
}d^{X} \Phi _{+}^{\prime \prime } \ dr \tag{A2}
\]%
and
\[
d_{2p,2p}^{X}=\int_{0}^{\infty }\sin^2 {\Theta} \ \Phi _{+}^{\prime
}d^{X} \Phi _{+}^{\prime \prime } \ dr \, . \tag{A3}
\]%
Similar expressions are obtained for transitions between states of
the second type. As a result, the TDM for the
strong transitions at large $J$ is proportional to
\[
\text{TDM} \propto d_{1p,1p}^{X} + d_{2p,2p}^{X}=\int_{0}^{\infty }\Phi
_{\pm}^{\prime }d^{X} \Phi _{\pm}^{\prime \prime } \ dr \, .
\]%

\begin{figure}[htbp]
    \centering
    \includegraphics[scale=0.2]{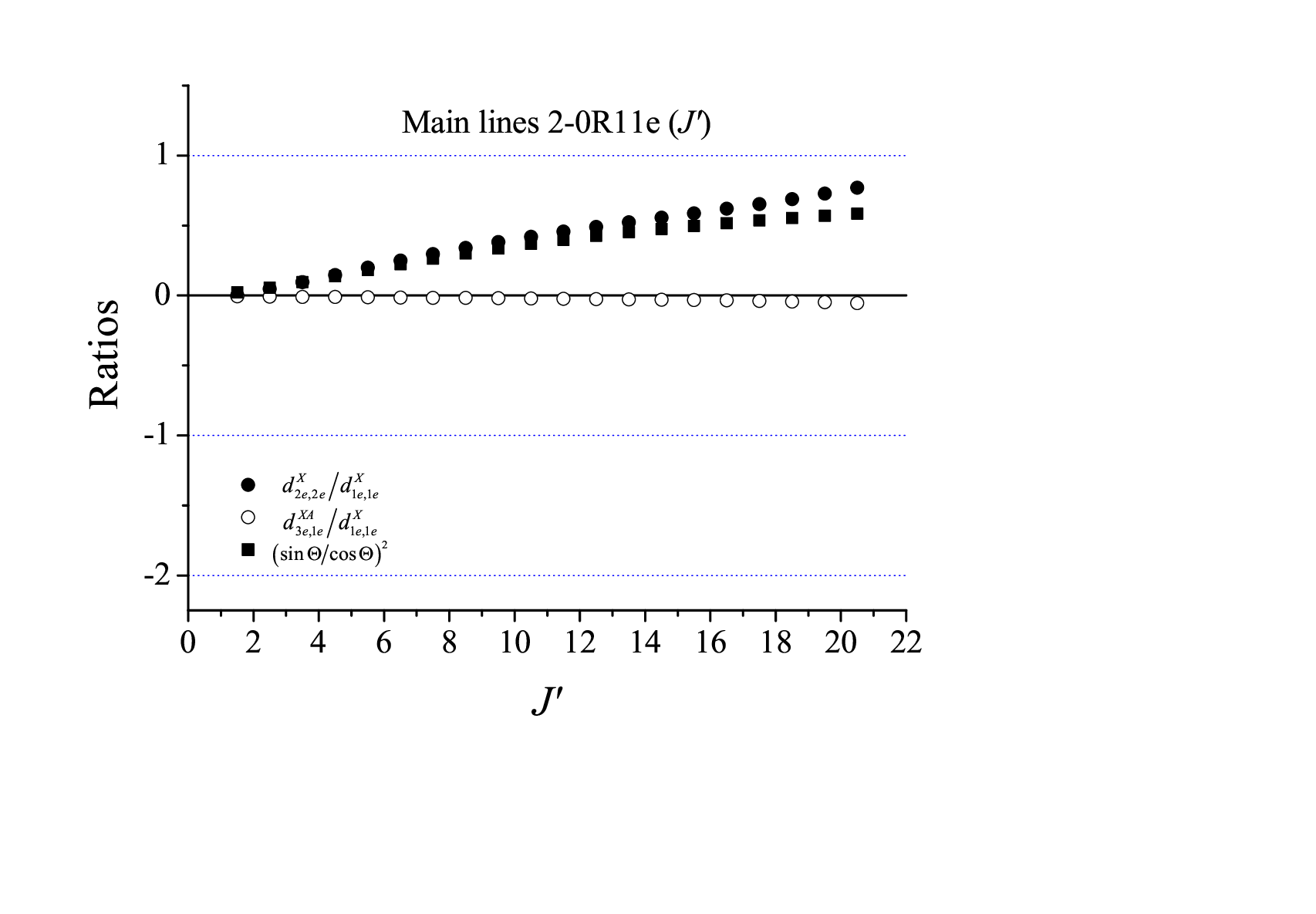}
    \includegraphics[scale=0.2]{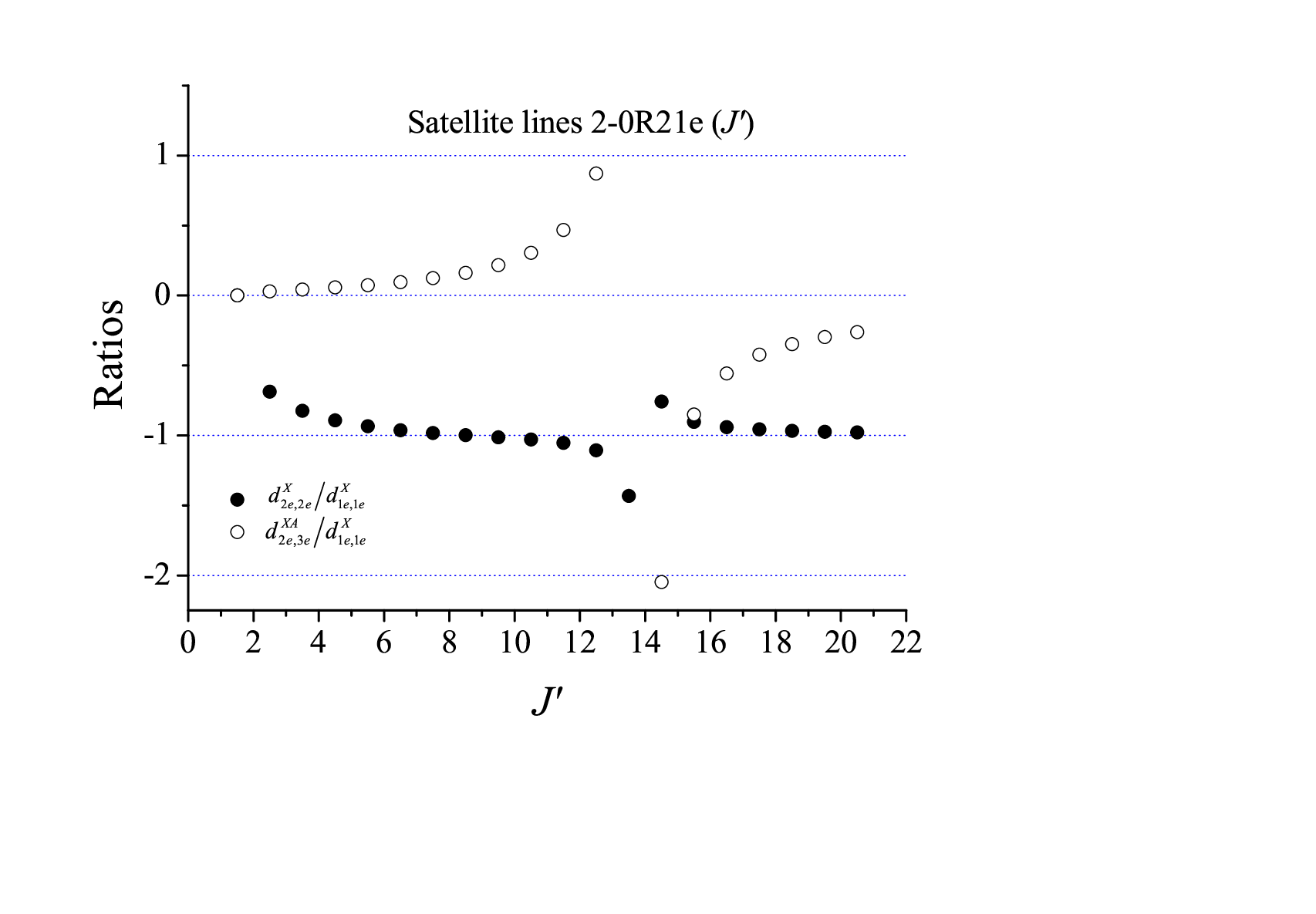}
    \includegraphics[scale=0.2]{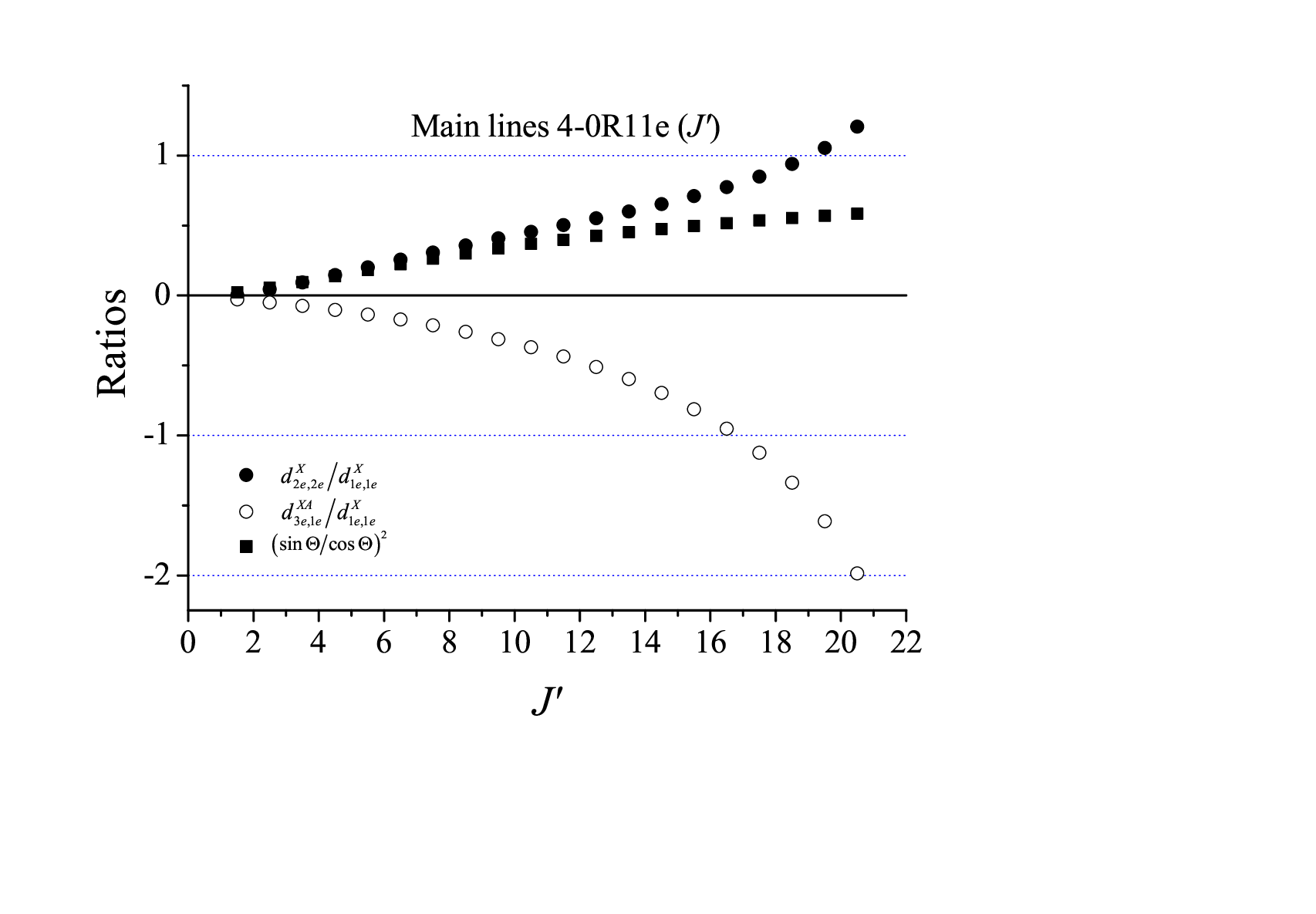}
    \includegraphics[scale=0.2]{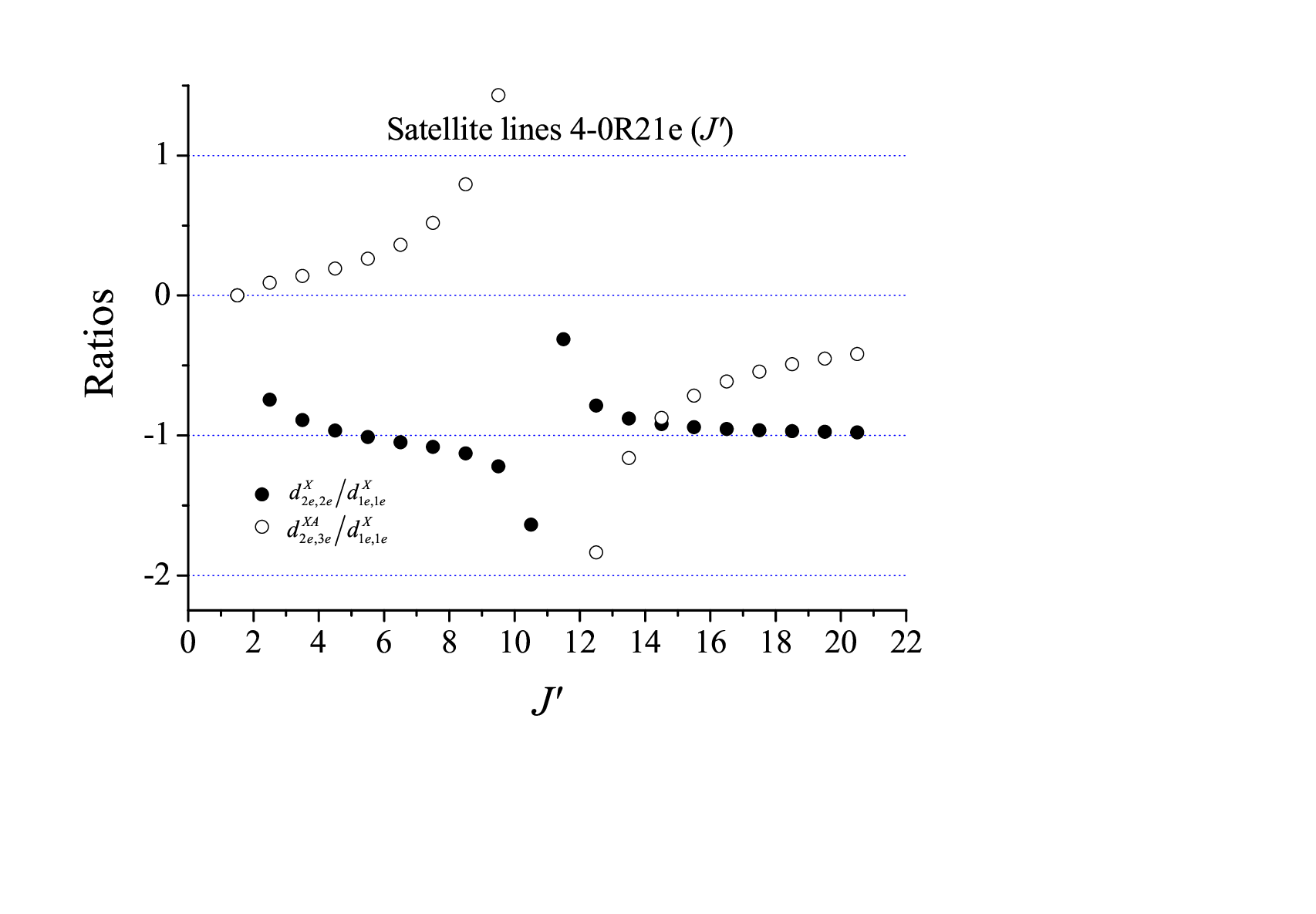}
    \caption{Ratios (A1) and (A4) for the main and satellite R-branch lines.}
    \label{figR}
\end{figure}

The above findings are confirmed by numerical calculations for the
general problem including interactions with the excited electronic state,
$A^2 \Sigma ^+$. This is illustrated in Fig. \ref{figR}, where ratios of various TDM components are shown for the $e\rightarrow e$ R-branch main and satellite transitions in the 2-0 and 4-0 bands. Consider them in turn.

In the right panels, ratio (A1) is plotted as function of $J^\prime$ for the satellite transitions (full circles). It is seen that Eq. (A1) is approximately obeyed except for the anomalies\footnote{Both $d_{1p,1p}^{X}$ and $d_{2p,2p}^{X}$ cross zero and change sign.} at $J^\prime=13.5$ and 10.5 for bands 2-0 and 4-0, respectively. In order to estimate the effect of the excited $A$ state, we also plotted the ratio
$d_{3p,3p}^{A}/d_{1p,1p}^{X}$ (open circles). It is seen that the $A$ state contributes significantly to the intensity of the satellite lines at $J^\prime\ge8.5$ in the 2-0 band and at all $J^\prime$ in the 4-0 band.

For comparison, the left panels show similar data for the main lines. Taking into account that $\Theta$ is a very slow function of $r$, we obtain the following approximate relation from Eqs. (A2) and (A3):
\[
\frac{d_{2p,2p}^X}{d_{1p,1p}^X} = \left(\frac{\sin{\Theta}}{\cos{\Theta}}\right)^2. \tag{A4}
\]
The left- and right-hand ratios are shown by filled circles and squares, respectively. It is seen that Eq. (A4) is approximately fulfilled except for high $J^\prime$ where the effect of the $A$ state increases. In the 2-0 band where the effect of the $A$ state is very small at all $J^\prime$, the breakdown of Eq. (A4) is also small. 

For the $Q$ branch, $d_{1p,1p}^{X}$ and $d_{2p,2p}^{X}$
contribute into the TDM in the form of $3d_{1p,1p}^{X}+d_{2p,2p}^{X}$. Thus, cancellation at high $J^\prime$ is incomplete. 
Moreover, analysis shows that the contributions of the $A$ state into TDMs cannot be neglected  at large $\Delta v$ and $J^\prime$,
seriously diminishing the intensity difference between the ``strong" and
``weak" lines in the $Q$ branch.

\begin{figure}[htbp]
    \centering
    \includegraphics[scale=0.2]{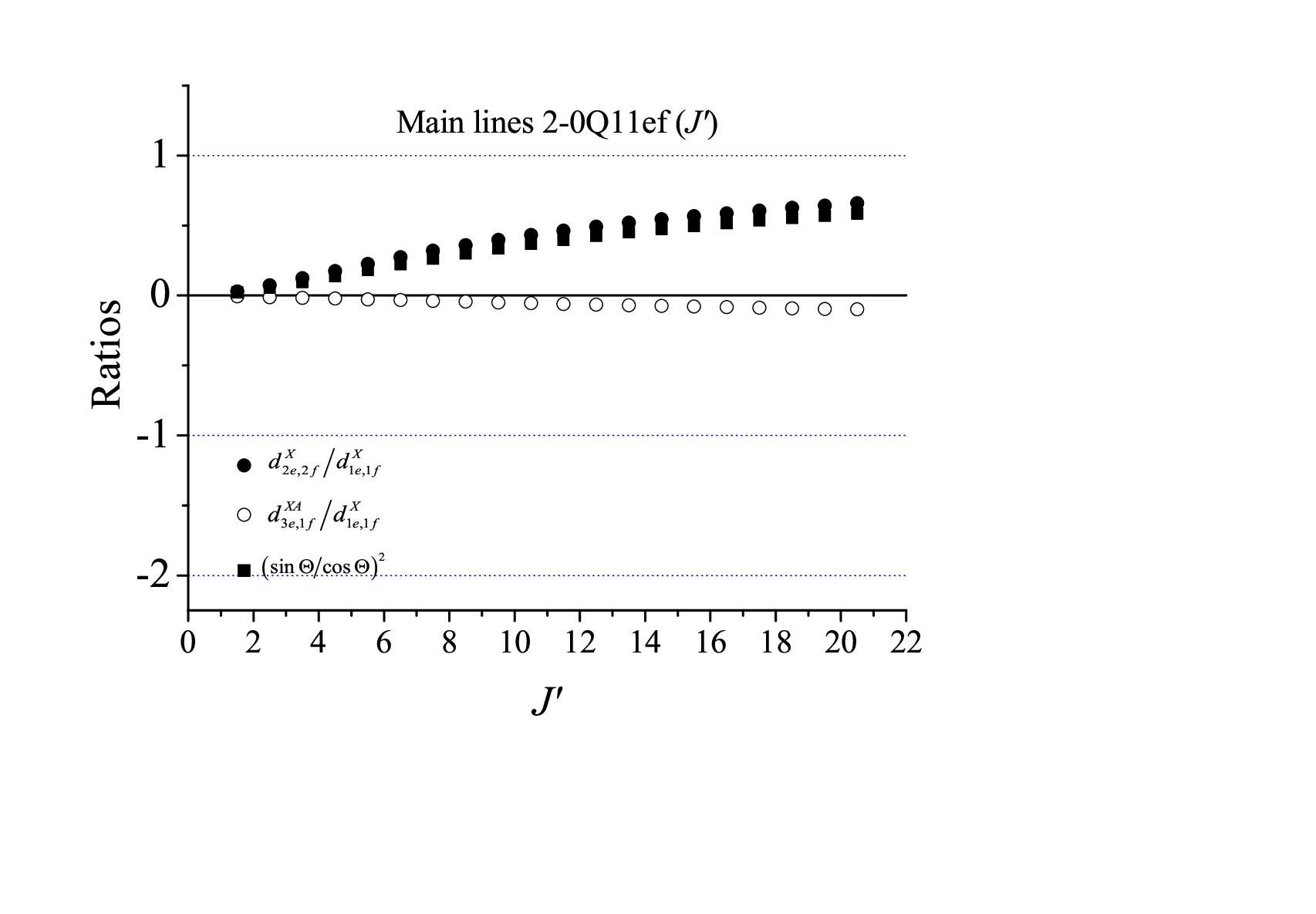}
    \includegraphics[scale=0.2]{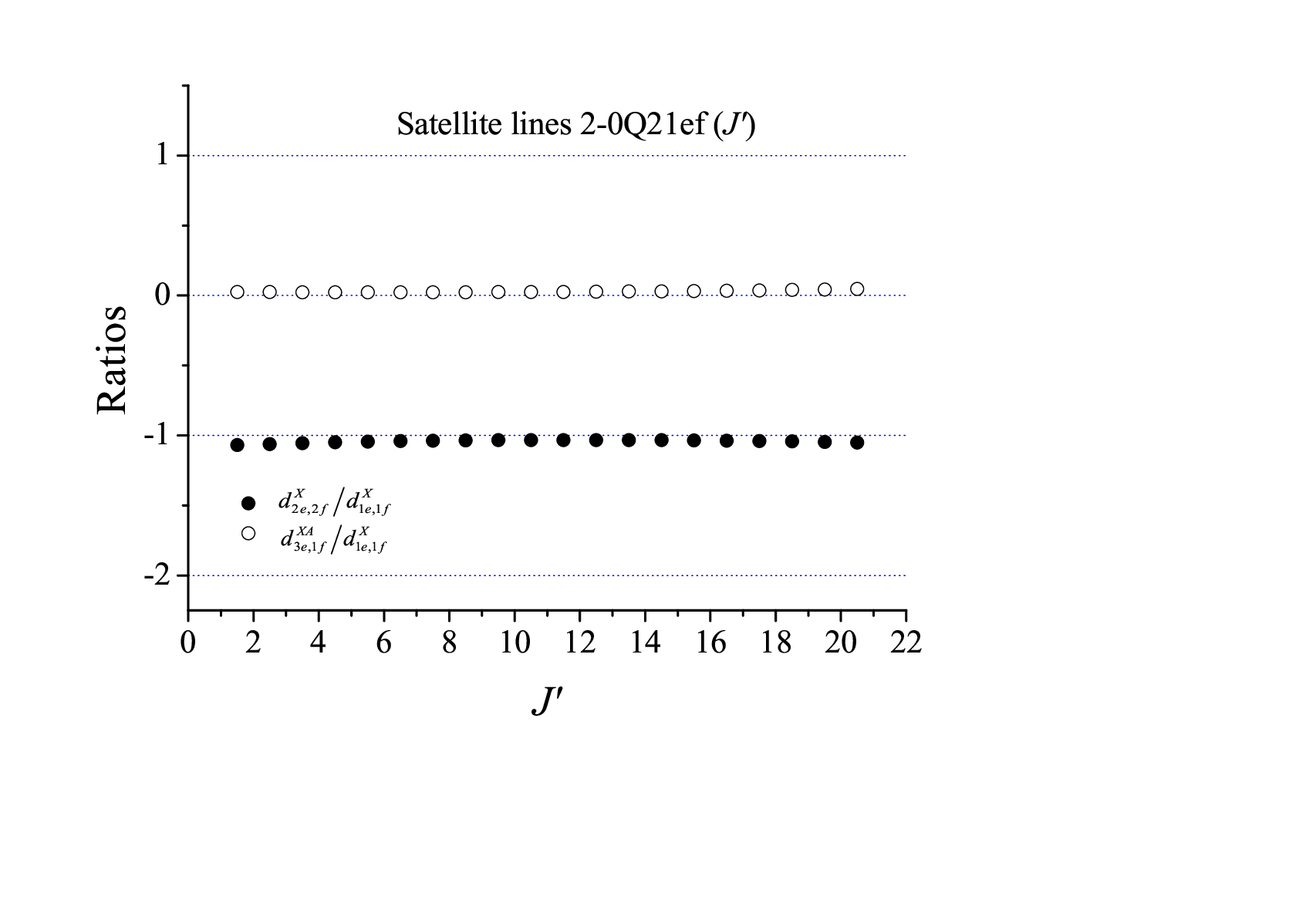}
    \includegraphics[scale=0.2]{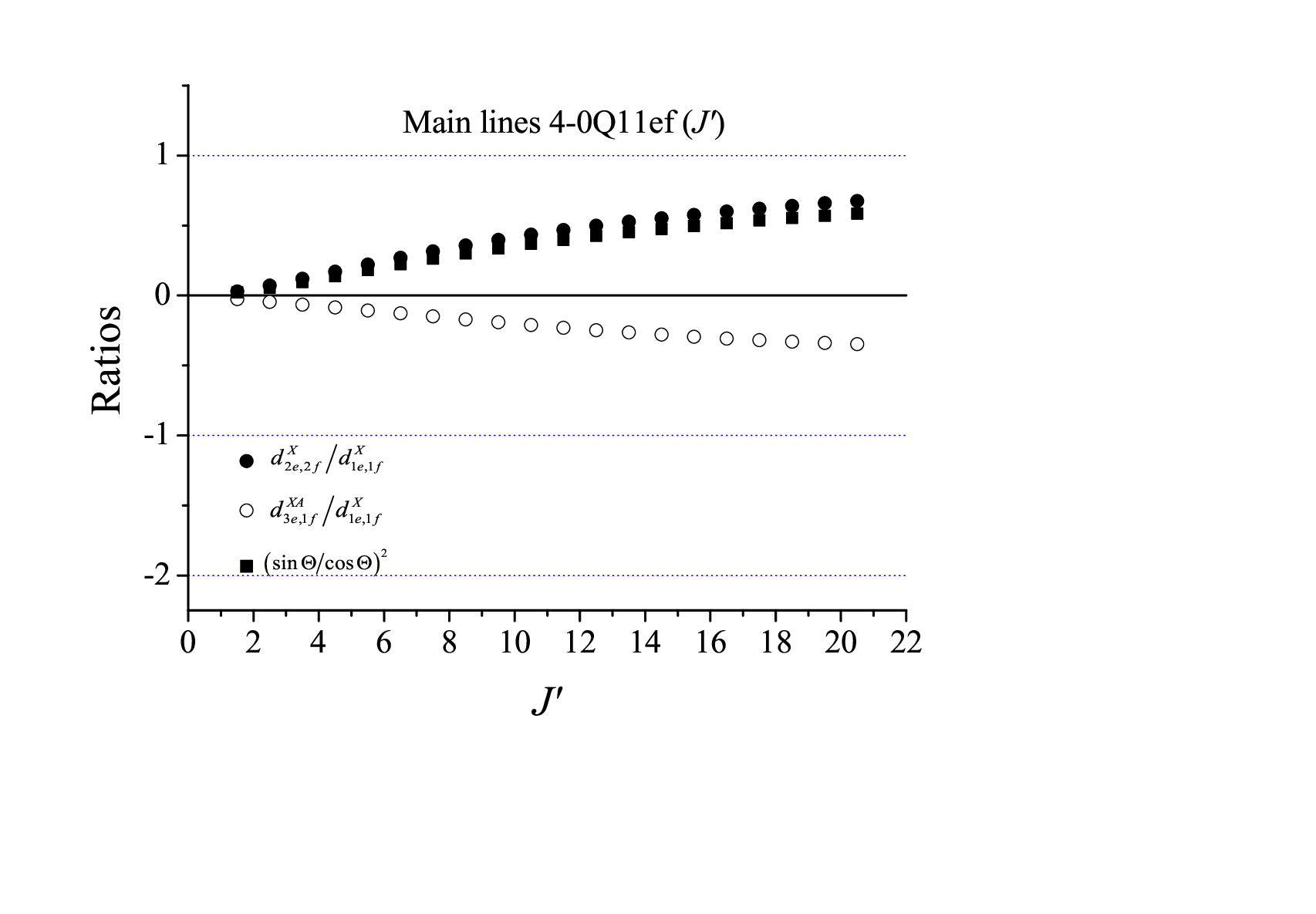}
    \includegraphics[scale=0.2]{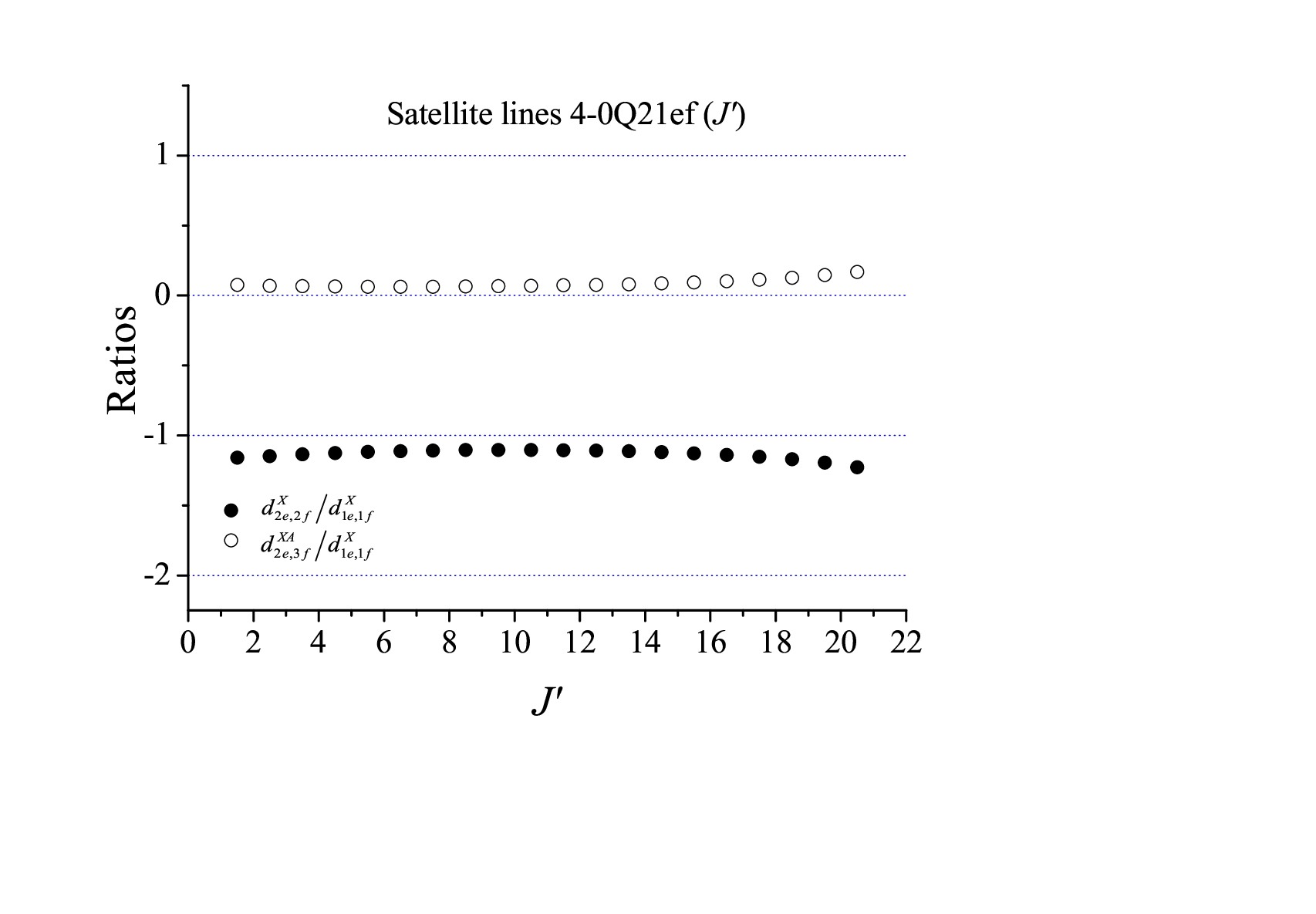}
    \caption{Ratios (A1) and (A4) for the main and satellite Q-branch lines.}
    \label{figQ}
\end{figure}

Figure \ref{figQ} shows the ratios for the Q lines. For the satellite lines, the effect of the $A$ state is small and Eq. (A1) is fulfilled at all $J^\prime$. For the main lines, the effect of the $A$ state is also small and Eq. (A4) is fulfilled at low $\Delta v$ and $J^\prime$, but becomes more important for the 4-0 band and $J^\prime\ge8.5$. 

To avoid confusion caused by the similarity of the right panels of Figs. \ref{figR} and \ref{figQ}, we should emphasize that the largest contribution to the TDM of the R lines comes from $d_{1p,1p}^{X}+d_{2p,2p}^{X}$, which tends to zero due to cancellaton expressed by Eq. (A1), whereas the largest contribution for the Q lines, $3d_{1p,1p}^{X}+d_{2p,2p}^{X}$, tends to $2d_{1p,1p}^{X}$ due to partial cancellation expressed  by the same Eq. (A1).

\bibliography{Paper_on_OH_2023}
\bibliographystyle{elsarticle-num}

\end{document}